\RequirePackage{fixltx2e}
\documentclass[a4paper]{isp}
\usepackage[american]{babel}
\usepackage{graphicx}
\usepackage{amsmath,amsthm,mathtools}
\usepackage{url}
\usepackage[capitalise,nameinlink]{cleveref}
\usepackage{xcolor}
\def\aap{A\&A\,  }

\def\apj{ApJ\,  }
\def\apjl{ApJ Letters,  }
\def\apjs{ApJS  }
\def\araa{ARA\&A  }

\def\mnras{MNRAS\,  }
\def\pasj{PASJ\,  }

\def\rmp{Rev. Mod. Phys.  }

\def\snr1993j{SN\,1993J~}

\begin{document}
\pdfgentounicode=1
\title
{
Energy conservation in the thin  layer approximation:
VI. Bubbles and super-bubbles 
}
\author{Lorenzo Zaninetti}
\institute{
Physics Department,
 via P. Giuria 1, I-10125 Turin, Italy \\
 \email{zaninetti@ph.unito.it}
}

\maketitle

\begin {abstract}
We model   the conservation of energy in the framework
of the thin layer approximation
for two types of interstellar medium (ISM).
In particular, we analyse  an ISM
in the presence of 
self-gravity
and a Gaussian  ISM  which  produces an asymmetry 
in the advancing shell.
The  astrophysical targets  to be simulated are 
the Fermi bubbles, 
the  local bubble,  and 
the  W4 super-bubble.
The  theory   of images is applied to a piriform curve,
which allows  deriving some  analytical  
formulae  for  the  observed  intensity  
in the case  of an optically thin medium.
\end{abstract}
\vspace{2pc}

\section{Introduction}

We now summarize 
the first uses  of some   words:
`super-shell'   can be found in  \cite{heiles1979},
where eleven H\,I objects were examined,
`super-bubble' 
can be found in  \cite{Cash1980}, 
where an X-ray region with a diameter of 450 pc
connected with Cyg X-6 and Cyg X-7 was observed 
and  
`worms', meaning  gas filaments crawling away from the galactic plane in
the inner Galaxy, can be found in \cite{Heiles1984}.
Super-bubbles   or super-shells  
can be  defined as cavities
with diameters greater than 100 pc and 
density  of matter lower 
than that of the surrounding interstellar medium (ISM)
\cite{Tenorio-Tagle1988}.
Bubbles have smaller diameters, 
between 10 pc and $100$ pc \cite{McCray1987}.
Some  models 
which   explain  super-shells as being due to 
the combined  explosions  of supernova in a cluster  of
massive stars will now be reviewed.
In  semi-analytical calculations,
the thin-shell approximation
can be the key to obtaining
the expansion of
the super-bubble; see,  for example,
\cite{McCray1987,mccrayapj87,MacLow1988,Igumenshchev1990,Basu1999}.
The Kompaneyets  approximation, 
see \cite{Kompaneets1960,Olano2009},
has  been used in order 
to model   the  super-bubble W4 \cite{Basu1999}
and the Orion–Eridanus super-bubble \cite{Pon2014,Pon2016}.
The hydro-dynamical approximation, 
with the inclusion of interstellar density gradients, 
can  produce  a blowout  into the galactic halo,
see  \cite{MacLow1989,Melioli2009}.
Recent Planck 353-GHz polarization observations
allow mapping the magnetic field,
see  \cite{Soler2018} for the Orion--Eridanus super-bubble,
and we recall that  
the expansion of super-bubbles in the presence of magnetic fields
has  been implemented in various  magneto-hydrodynamic 
codes, see \cite{Tomisaka1992,Rafikov2000}.
The present paper  
derives  
the  equation of motion
for two different  ISMs 
in the framework  of the energy conservation  
for the thin layer approximation,
see Section  \ref{section_motion},
compares 
the observed and the theoretical sections  
for Fermi bubbles, the local bubble, and the W4 super-bubble,
see Section \ref{section_astrophysical}, 
and derives  
a new analytical formula for the
theoretical profile in intensity 
using  the  piriform curve,  
see Section \ref{section_image}.

\section{The equations of motion}

\label{section_motion}
We start with 
the conservation of kinetic energy in
spherical coordinates
in  the framework of the thin
layer approximation  
\begin{equation}
\frac{1}{2} M_0(r_0) \,v_0^2 = \frac{1}{2}M(r) \,v^2 
\quad ,
\label{cons_rel_energy}
\end{equation}
where  $M_0(r_0)$ and $M(r)$ are the swept masses at $r_0$ and $r$,
while $v_0$ and $v$ are the velocities of the thin layer at $r_0$ and $r$.
The above equation holds for  the  solid angle  $\Delta \Omega$,
which in the following is unity. 
We now present two asymmetric equations of motion for 
bubbles and super-bubbles.
The above equation is  a differential equation 
of the first   order:
\begin{equation}
\frac{1}{2} M_0(r_0) \,v_0^2 = \frac{1}{2}M(r) \,(\frac{dr}{dt})^2 
\quad .
\label{differential_equation}
\end{equation}
The asymmetry is due to a gradient  of the number of particles  with  the
distance or galactic height, $z$, which is parametrized as
\begin{equation}
n(z)  =
n_1 e^{- z^2 /{H_1}^2}+
n_2 e^{- z^2 /{H_2}^2}+
n_3 e^{-  | z |  /{H_3}}
\,.
\label{equation:ism}
\end{equation}
where 
$n_1$=0.395 ${\mathrm{particles~}}{\mathrm{cm}^{-3}}$, $H_1$=127
        \mbox{pc},
        $n_2$=0.107 $\mathrm{particles~}{\mathrm{cm}^{-3}}$, $H_2$=318
        \mbox{pc},
        $n_3$=0.064 $\mathrm{particles~}{\mathrm{cm}^{-3}}$, and  $H_3$=403
        \mbox{pc}
\cite{Bisnovatyi1995,Dickey1990,Lockman1984}.
In the framework of Cartesian coordinates, $(x,y,z)$,
when the explosion starts at $(0,0,0)$ 
we have an up--down symmetry,  
$r(x,y,-z)=r(x,y,z)$
and a 
right--left symmetry $r(x,-y,z)=r(x,y,z)$.
Conversely, when the explosion starts 
at $(0,0,z_{OB})$,    where  $z_{OB}$ 
represents  the distance in pc  from the  position of  the
OB association which generate the phenomena,
we have only  
a 
right--left symmetry $r(x,-y,z)=r(x,y,z)$.
  
\subsection{Numerical  methods}

In the absence of an analytical 
solution for the  trajectory, we
outline  four  ways  which   allow 
obtaining a numerical solution.
\begin{enumerate}
\item 
Evaluation of the   numerical solution 
with the the Runge–Kutta method.
\item 
A non-linear method 
which obtains  the trajectory  by the
following non-linear equation 
\begin{equation}
\int_{r_0}^r \frac{1}{(\frac{dr}{dt})} dr  = t-t_0
\quad .
\end{equation}
\item 
The   Euler method,  
which solves the following recursive equations  
\begin{subequations}
\begin{align}
r_{n+1} = r_n + v_n \Delta t
\label{recursive1}   
 \\
v_{n+1} = v_n 
\Bigl (\frac {M_n(r_n)}{M_{n+1} (r_{n+1})} \Bigr )^{1/2}
\label{velocityeuler}
\quad,
\end{align}
\end{subequations}
where  $r_n$, $v_n$, and $M_n$ are the temporary radius,
velocity,  and total mass, respectively,
$\Delta t $ is the time step,  and $n$ is the index.
\item 
 A  power series solution of the form
\begin{equation}
r(t) = a_0 +a_1  (t-t_0) +a_2  (t-t_0)^2+a_3  (t-t_0)^3 + \dots
\quad ,
\label{rtseries}
\end{equation}
see  \cite{Tenenbaum1963,Ince2012}.
\end{enumerate}

The  case  of an expansion  that starts  from a given 
galactic height $z$,  denoted by $z_{\mathrm{OB}}$,
which  represents  the OB associations,  
is  also analysed.
The advancing expansion is computed in a 3D Cartesian
coordinate system ($x,y,z$)  with the centre 
of the explosion at  (0,0,0).
The explosion is better visualized  
in a 3D Cartesian
coordinate system ($X,Y,Z$) in which the galactic plane
is given by $Z=0$.
The following 
translation, $T_{\mathrm{OB}}$,   
relates  the two Cartesian coordinate  systems 
\begin{equation}
T_{\mathrm{OB}} ~
 \left\{ 
  \begin {array}{l} 
  X=x  \\\noalign{\medskip}
  Y=y  \\\noalign{\medskip}
  Z=z+ z_{\mathrm{OB}}
  \end {array} 
  \right.  \quad , 
\label{ttranslation}
\end{equation}
where $z_{\mathrm{OB}}$  
is the distance  in pc  of the 
OB associations   from the galactic plane.
In the case of $z_{OB} \neq 0$, the  two masses
which appear in Eq.~(\ref{velocityeuler})
should be carefully evaluated.

\subsection{Medium in the presence of self-gravity}

\begin{figure*}
\begin{center}
\includegraphics[width=6cm]{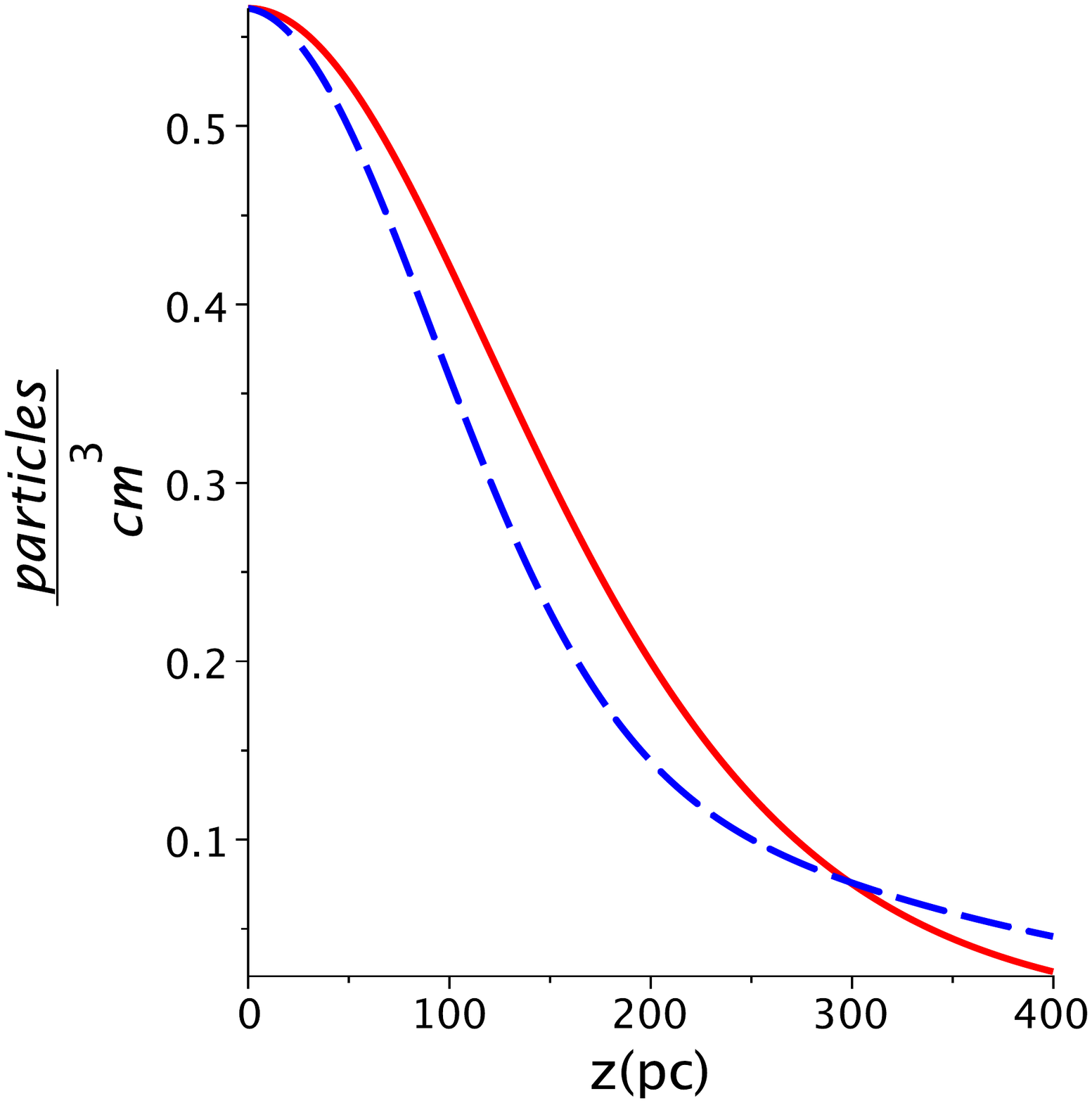}
\end {center}
\caption
{
Profiles of density versus scale height $z$:
the   disk  in presence of 
self-gravity  
as
given by Eq.~(\ref{sech2})
when $h=90$\ pc
(dashed blue line)
and
the
three-component exponential distribution
as
given by Eq.~(\ref{equation:ism})
(red full line).
}
\label{ism_sech2}
    \end{figure*}

We assume that 
the  number density
distribution scales as   
\begin{equation}
n(z) = n_0 sech^2 (\frac{z}{2\,h})
\quad ,
\label{sech2}
\end{equation}
where $n_0$ is the density at $z=0$,
$h$ is a scaling parameter, 
and  sech is the hyperbolic secant 
\cite{Spitzer1942,Rohlfs1977,Bertin2000,Padmanabhan_III_2002}.
In order  to include the boundary conditions 
we  assume that the density of the medium 
around the OB associations 
scales with the  
self-gravity
piece-wise dependence
\begin{equation}
 \rho (r;r_0)  =\Bigg \{ \begin{array}{ll}
            \rho_c                      & \mbox {if $r \leq r_0 $ } \\
            \rho_c\, sech^2 (\frac{z}{2\,h})    & \mbox {if $r >     r_0 $}
            \end{array}
\label{piecewisesech2}
\quad ,
\end{equation}
where  $\rho_c$  is the density at $z=0$.
In order to find an acceptable  value  of $h$,
we make a comparison  with 
Eq.~(\ref{equation:ism}), 
after which we choose $h=90$\ pc, 
see Figure \ref{ism_sech2}.

\begin{figure*}
\begin{center}
\includegraphics[width=6cm]{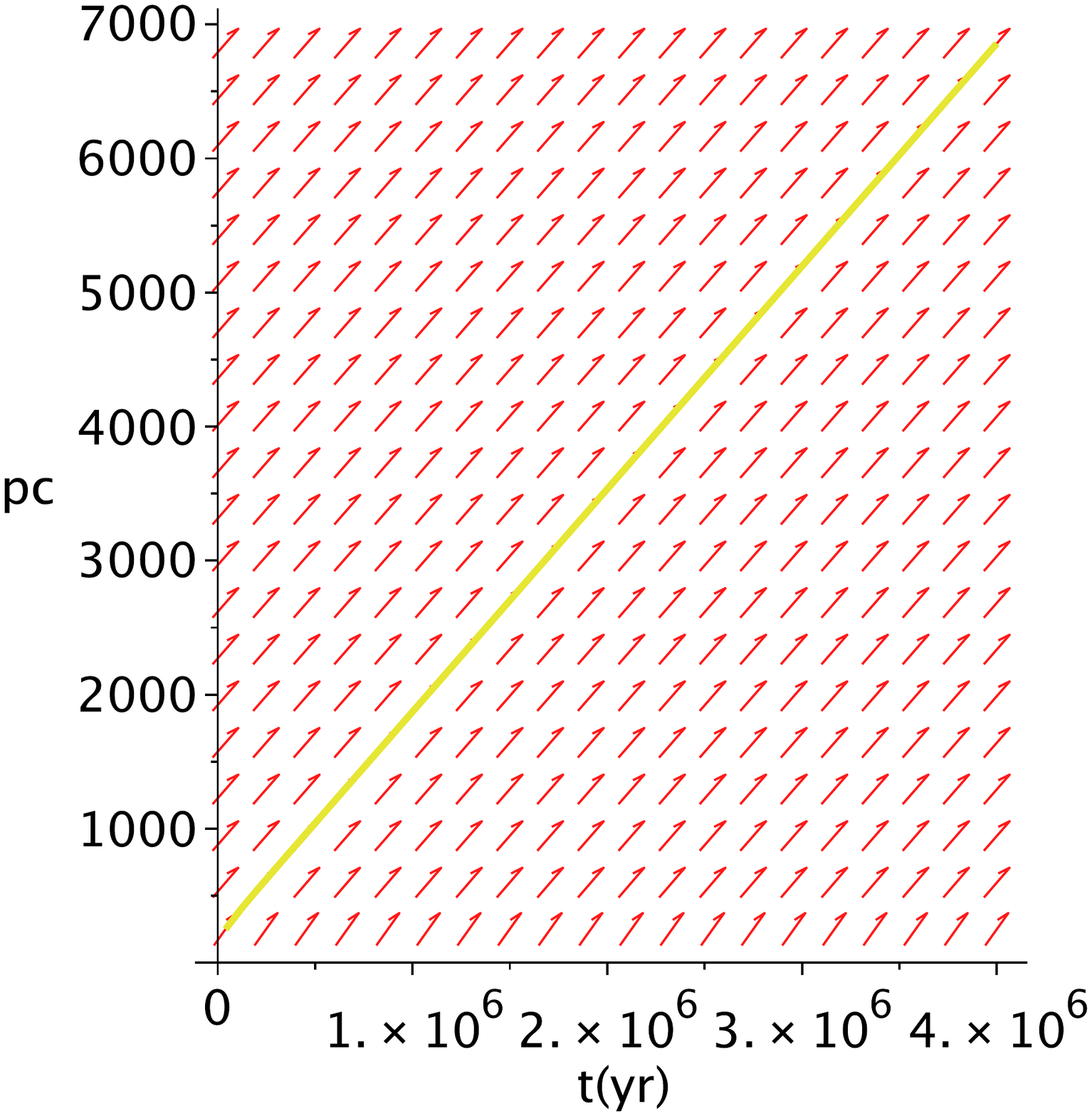}
\end {center}
\caption
{
Phase plane portrait and numerical solution (full yellow line)
for a  medium  in the presence of 
self-gravity
as
given by Eq.~(\ref{sech2})
when  
$r_0=250$\ pc,
$h=90   $\ pc,
$t=4\,10^6$\ yr,
$t_0 =4 \,10^4$ yr  and
$v_0\,=2000$ km s$^{-1}$.
}
\label{sech2_portrait}
    \end{figure*}

\begin{figure*}
\begin{center}
\includegraphics[width=6cm]{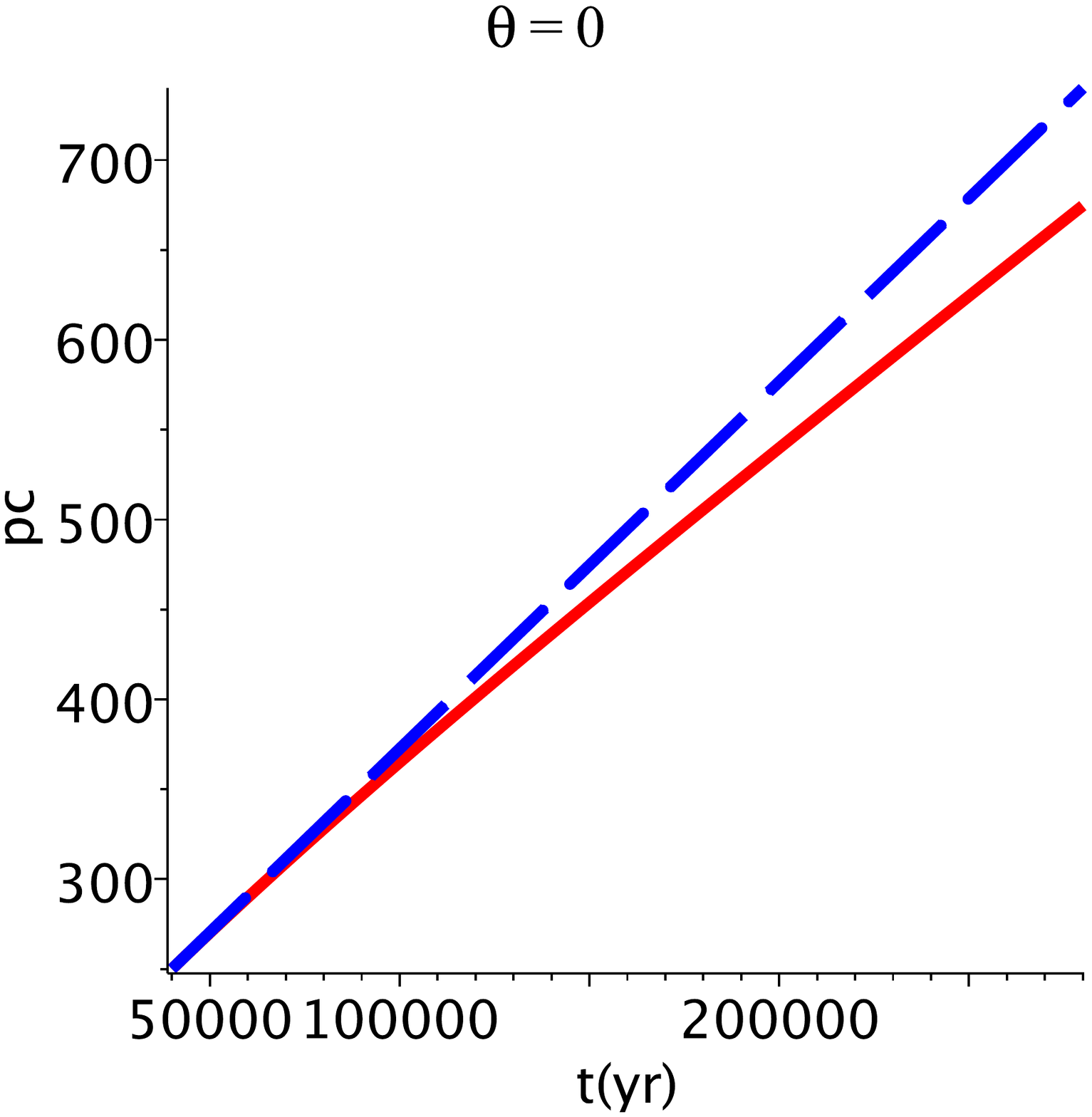}
\end {center}
\caption
{
Numerical solution (red full line)    
and Taylor expansion of the solution (dashed blue line),
parameters as in Figure \ref{sech2_portrait}
but $t=2.8\,10^5$\ yr, 
}
\label{taylor_num_sech2}
    \end{figure*}

The mass $M_0$
swept in the interval [0,$r_0$]
is
\begin{equation}
M_0(\rho_c,r_0) =
\frac{4}{3}\,\rho_{{c}}\pi \,{r_{{0}}}^{3}
\quad.
\nonumber
\end{equation}
The total mass $ M(r;r_0,\rho_c,h)$
swept in the interval [0,r]
is
\begin{eqnarray}
M (r;r_0,\rho_c,h)=
{\frac {\rho_{{c}}{r_{{0}}}^{3}}{3}}-4\,{\frac {\rho_{{c}}{r}^{2}h}{
\cos \left( \theta \right) } \left( 1+{{\rm e}^{{\frac {r\cos \left( 
\theta \right) }{h}}}} \right) ^{-1}}+4\,{\frac {\rho_{{c}}{r}^{2}h}{
\cos \left( \theta \right) }}-8\,{\frac {\rho_{{c}}{h}^{2}r}{ \left( 
\cos \left( \theta \right)  \right) ^{2}}\ln  \left( 1+{{\rm e}^{{
\frac {r\cos \left( \theta \right) }{h}}}} \right) }
\nonumber \\
-8\,{\frac {\rho_{
{c}}{h}^{3}}{ \left( \cos \left( \theta \right)  \right) ^{3}}{\it 
polylog} \left( 2,-{{\rm e}^{{\frac {r\cos \left( \theta \right) }{h}}
}} \right) }+4\,{\frac {\rho_{{c}}{r_{{0}}}^{2}h}{\cos \left( \theta
 \right) } \left( 1+{{\rm e}^{{\frac {r_{{0}}\cos \left( \theta
 \right) }{h}}}} \right) ^{-1}}
\nonumber \\
-4\,{\frac {\rho_{{c}}{r_{{0}}}^{2}h}{
\cos \left( \theta \right) }}+8\,{\frac {\rho_{{c}}{h}^{2}r_{{0}}}{
 \left( \cos \left( \theta \right)  \right) ^{2}}\ln  \left( 1+{
{\rm e}^{{\frac {r_{{0}}\cos \left( \theta \right) }{h}}}} \right) }+8
\,{\frac {\rho_{{c}}{h}^{3}}{ \left( \cos \left( \theta \right) 
 \right) ^{3}}{\it polylog} \left( 2,-{{\rm e}^{{\frac {r_{{0}}\cos
 \left( \theta \right) }{h}}}} \right) }
\end{eqnarray}
where $\theta$ is the polar angle 
and
the polylog operator  is
defined by
\begin{equation}
polylog(s,z) =
\mathrm{Li}_{s}\left(z\right) =\sum_{n=1}^{\infty}\frac{z^{n}}{n^{s}}
\quad 
\end{equation}
where   $\mathrm{Li}_{s}\left(z\right)$ is
the Dirichlet series.
The positive solution of Eq.~(\ref{cons_rel_energy})
gives the velocity as a function 
of the radius:
\begin{equation}
v(r;r_0,v_0,h) =
\frac{AN}{AD}
\quad  ,
\end{equation}
where
\begin{eqnarray}
AN=
-\Big ( {-24\,\cos   ( \theta ) r_{{0}}   ( {{\rm e}^{{
\frac {\cos   ( \theta )    ( r+r_{{0}} ) }{h}}}}+{
{\rm e}^{{\frac {\cos   ( \theta ) r_{{0}}}{h}}}}+{{\rm e}^{{
\frac {r\cos   ( \theta ) }{h}}}}+1 )    ( -{h}^{3}
   ( {{\rm e}^{{\frac {\cos   ( \theta )    ( r+r_{{0}}
 ) }{h}}}}+{{\rm e}^{{\frac {\cos   ( \theta ) r_{{0}}
}{h}}}}+{{\rm e}^{{\frac {r\cos   ( \theta ) }{h}}}}
}
\nonumber \\
+1
 ) 
{
{\it polylog}   ( 2,-{{\rm e}^{{\frac {\cos   ( \theta
 ) r_{{0}}}{h}}}} ) +{h}^{3}   ( {{\rm e}^{{\frac {\cos
   ( \theta )    ( r+r_{{0}} ) }{h}}}}+{{\rm e}^{{
\frac {\cos   ( \theta ) r_{{0}}}{h}}}}+{{\rm e}^{{\frac {r
\cos   ( \theta ) }{h}}}}+1 ) {\it polylog}   ( 2,-{
{\rm e}^{{\frac {r\cos   ( \theta ) }{h}}}} ) 
}
\nonumber \\
{
+\cos
   ( \theta )    (    ( -{h}^{2}r_{{0}}\ln    ( 1+{
{\rm e}^{{\frac {\cos   ( \theta ) r_{{0}}}{h}}}} ) +{h
}^{2}r\ln    ( 1+{{\rm e}^{{\frac {r\cos   ( \theta ) }{h}
}}} ) -1/2\,\cos   ( \theta )    ( 1/12\,{r_{{0}}}^{
3}\cos   ( \theta ) +h{r}^{2}-h{r_{{0}}}^{2} ) 
 ) 
}
\nonumber \\
{
{{\rm e}^{{\frac {\cos   ( \theta )    ( r+r_{{0}
} ) }{h}}}}-{h}^{2}r_{{0}}   ( {{\rm e}^{{\frac {\cos   ( 
\theta ) r_{{0}}}{h}}}}+{{\rm e}^{{\frac {r\cos   ( \theta
 ) }{h}}}}+1 ) \ln    ( 1+{{\rm e}^{{\frac {\cos
   ( \theta ) r_{{0}}}{h}}}} ) +{h}^{2}r   ( {
{\rm e}^{{\frac {\cos   ( \theta ) r_{{0}}}{h}}}}+{{\rm e}^{{
\frac {r\cos   ( \theta ) }{h}}}}+1 ) 
\times 
}
\nonumber \\
{
\ln    ( 1+{
{\rm e}^{{\frac {r\cos   ( \theta ) }{h}}}} ) 
-1/2\,
\cos   ( \theta )    (    ( 1/12\,{r_{{0}}}^{3}\cos
   ( \theta ) -h{r_{{0}}}^{2} ) {{\rm e}^{{\frac {\cos
   ( \theta ) r_{{0}}}{h}}}}
}
\nonumber \\
{
+   ( 1/12\,{r_{{0}}}^{3}\cos
   ( \theta ) +h{r}^{2} ) {{\rm e}^{{\frac {r\cos
   ( \theta ) }{h}}}}+1/12\,{r_{{0}}}^{3}\cos   ( \theta
 )  )  )  ) }\Big )^{1/2}
 \cos   ( \theta ) {\it 
v_0}\,r_{{0}}
\end{eqnarray}
and  
\begin{eqnarray}
AD=
-{r_{{0}}}^{3}{{\rm e}^{{\frac {\cos \left( \theta \right)  \left( r+r
_{{0}} \right) }{h}}}} \left( \cos \left( \theta \right)  \right) ^{3}
+24\,{h}^{2}r{{\rm e}^{{\frac {\cos \left( \theta \right)  \left( r+r_
{{0}} \right) }{h}}}}\ln  \left( 1+{{\rm e}^{{\frac {r\cos \left( 
\theta \right) }{h}}}} \right) \cos \left( \theta \right) 
\nonumber \\
-24\,{h}^{2}
r_{{0}}{{\rm e}^{{\frac {\cos \left( \theta \right)  \left( r+r_{{0}}
 \right) }{h}}}}\ln  \left( 1+{{\rm e}^{{\frac {\cos \left( \theta
 \right) r_{{0}}}{h}}}} \right) \cos \left( \theta \right) -12\,h{r}^{
2}{{\rm e}^{{\frac {\cos \left( \theta \right)  \left( r+r_{{0}}
 \right) }{h}}}} \left( \cos \left( \theta \right)  \right) ^{2}
\nonumber \\
+12\,h
{r_{{0}}}^{2}{{\rm e}^{{\frac {\cos \left( \theta \right)  \left( r+r_
{{0}} \right) }{h}}}} \left( \cos \left( \theta \right)  \right) ^{2}-
{r_{{0}}}^{3}{{\rm e}^{{\frac {r\cos \left( \theta \right) }{h}}}}
 \left( \cos \left( \theta \right)  \right) ^{3}-{r_{{0}}}^{3}{{\rm e}
^{{\frac {\cos \left( \theta \right) r_{{0}}}{h}}}} \left( \cos
 \left( \theta \right)  \right) ^{3}
\nonumber \\
+24\,{h}^{3}{{\rm e}^{{\frac {\cos
 \left( \theta \right)  \left( r+r_{{0}} \right) }{h}}}}{\it polylog}
 \left( 2,-{{\rm e}^{{\frac {r\cos \left( \theta \right) }{h}}}}
 \right) -24\,{h}^{3}{{\rm e}^{{\frac {\cos \left( \theta \right) 
 \left( r+r_{{0}} \right) }{h}}}}{\it polylog} \left( 2,-{{\rm e}^{{
\frac {\cos \left( \theta \right) r_{{0}}}{h}}}} \right) 
\nonumber \\
+24\,{h}^{2}r
{{\rm e}^{{\frac {r\cos \left( \theta \right) }{h}}}}\ln  \left( 1+{
{\rm e}^{{\frac {r\cos \left( \theta \right) }{h}}}} \right) \cos
 \left( \theta \right) +24\,{h}^{2}r{{\rm e}^{{\frac {\cos \left( 
\theta \right) r_{{0}}}{h}}}}\ln  \left( 1+{{\rm e}^{{\frac {r\cos
 \left( \theta \right) }{h}}}} \right) \cos \left( \theta \right) 
\nonumber \\
-24
\,{h}^{2}r_{{0}}{{\rm e}^{{\frac {r\cos \left( \theta \right) }{h}}}}
\ln  \left( 1+{{\rm e}^{{\frac {\cos \left( \theta \right) r_{{0}}}{h}
}}} \right) \cos \left( \theta \right) -24\,{h}^{2}r_{{0}}{{\rm e}^{{
\frac {\cos \left( \theta \right) r_{{0}}}{h}}}}\ln  \left( 1+{{\rm e}
^{{\frac {\cos \left( \theta \right) r_{{0}}}{h}}}} \right) \cos
 \left( \theta \right) 
\nonumber \\
-12\,h{r}^{2}{{\rm e}^{{\frac {r\cos \left( 
\theta \right) }{h}}}} \left( \cos \left( \theta \right)  \right) ^{2}
+12\,h{r_{{0}}}^{2}{{\rm e}^{{\frac {\cos \left( \theta \right) r_{{0}
}}{h}}}} \left( \cos \left( \theta \right)  \right) ^{2}-{r_{{0}}}^{3}
 \left( \cos \left( \theta \right)  \right) ^{3}
\nonumber \\
+24\,{h}^{3}{{\rm e}^{
{\frac {r\cos \left( \theta \right) }{h}}}}{\it polylog} \left( 2,-{
{\rm e}^{{\frac {r\cos \left( \theta \right) }{h}}}} \right) -24\,{h}^
{3}{{\rm e}^{{\frac {r\cos \left( \theta \right) }{h}}}}{\it polylog}
 \left( 2,-{{\rm e}^{{\frac {\cos \left( \theta \right) r_{{0}}}{h}}}}
 \right) 
\nonumber \\
+24\,{h}^{3}{{\rm e}^{{\frac {\cos \left( \theta \right) r_{{0
}}}{h}}}}{\it polylog} \left( 2,-{{\rm e}^{{\frac {r\cos \left( \theta
 \right) }{h}}}} \right) -24\,{h}^{3}{{\rm e}^{{\frac {\cos \left( 
\theta \right) r_{{0}}}{h}}}}{\it polylog} \left( 2,-{{\rm e}^{{\frac 
{\cos \left( \theta \right) r_{{0}}}{h}}}} \right) 
\nonumber \\
+24\,{h}^{2}r\ln 
 \left( 1+{{\rm e}^{{\frac {r\cos \left( \theta \right) }{h}}}}
 \right) \cos \left( \theta \right) -24\,{h}^{2}r_{{0}}\ln  \left( 1+{
{\rm e}^{{\frac {\cos \left( \theta \right) r_{{0}}}{h}}}} \right) 
\cos \left( \theta \right) 
\nonumber  \\
+24\,{h}^{3}{\it polylog} \left( 2,-{
{\rm e}^{{\frac {r\cos \left( \theta \right) }{h}}}} \right) -24\,{h}^
{3}{\it polylog} \left( 2,-{{\rm e}^{{\frac {\cos \left( \theta
 \right) r_{{0}}}{h}}}} \right) 
\end{eqnarray}
The differential equation which governs
the  motion for the medium
in the presence of 
self-gravity
is 
\begin{eqnarray}
 \Big ( {\frac {\rho_{{c}}{r_{{0}}}^{3}}{3}}-4\,{\frac {\rho_{{c}}
   ( r   ( t   )    ) ^{2}h}{\cos   ( \theta   ) 
}   ( 1+{{\rm e}^{{\frac {r   ( t   ) \cos   ( \theta
   ) }{h}}}}   ) ^{-1}}+4\,{\frac {\rho_{{c}}   ( r   ( 
t   )    ) ^{2}h}{\cos   ( \theta   ) }}
\nonumber  \\
-8\,{\frac {
\rho_{{c}}{h}^{2}r   ( t   ) }{   ( \cos   ( \theta
   )    ) ^{2}}\ln    ( 1+{{\rm e}^{{\frac {r   ( t
   ) \cos   ( \theta   ) }{h}}}}   ) }-8\,{\frac {\rho_
{{c}}{h}^{3}}{   ( \cos   ( \theta   )    ) ^{3}}{\it 
polylog}   ( 2,-{{\rm e}^{{\frac {r   ( t   ) \cos   ( 
\theta   ) }{h}}}}   ) }+4\,{\frac {\rho_{{c}}{r_{{0}}}^{2}h}{
\cos   ( \theta   ) }   ( 1+{{\rm e}^{{\frac {r_{{0}}\cos
   ( \theta   ) }{h}}}}   ) ^{-1}}
\nonumber \\
-4\,{\frac {\rho_{{c}}{r_
{{0}}}^{2}h}{\cos   ( \theta   ) }}+8\,{\frac {\rho_{{c}}{h}^{2
}r_{{0}}}{   ( \cos   ( \theta   )    ) ^{2}}\ln 
   ( 1+{{\rm e}^{{\frac {r_{{0}}\cos   ( \theta   ) }{h}}}}
   ) }+8\,{\frac {\rho_{{c}}{h}^{3}}{   ( \cos   ( \theta
   )    ) ^{3}}{\it polylog}   ( 2,-{{\rm e}^{{\frac {r_{{0
}}\cos   ( \theta   ) }{h}}}}   ) } \Big  )  \Big  ( {\frac 
{\rm d}{{\rm d}t}}r   ( t   )  \Big  ) ^{2}
\nonumber 
\\
-{\frac {\rho_{{c}}{
r_{{0}}}^{3}{v_{{0}}}^{2}}{3}}=0
\quad  ,
\label{eqndiffsech2}
\end{eqnarray}
and does not  have an analytical solution. 
Figure \ref{sech2_portrait}   shows  
the numerical solution obtained with the 
Runge–Kutta method.
A Taylor expansion of order 3 
of Eq.~(\ref{eqndiffsech2}) 
gives 
\begin{equation}
r(t;t_0,v_0,r_0,h)=
r_{{0}}+v_{{0}} \left( t-t_{{0}} \right) -3\,{\frac {{v_{{0}}}^{2}
 \left( t-t_{{0}} \right) ^{2}}{r_{{0}}}{{\rm e}^{{\frac {r_{{0}}\cos
 \left( \theta \right) }{h}}}} \left(  \left( {{\rm e}^{{\frac {r_{{0}
}\cos \left( \theta \right) }{h}}}} \right) ^{2}+2\,{{\rm e}^{{\frac {
r_{{0}}\cos \left( \theta \right) }{h}}}}+1 \right) ^{-1}}
\quad ,
\end{equation}
and  Figure \ref{taylor_num_sech2}   shows  
the numerical solution obtained by the 
Runge–Kutta method
and  the series solution
up to a time for which the percentage error  is less than $10\%$.
 
\subsection{Gaussian medium}

\begin{figure*}
\begin{center}
\includegraphics[width=6cm]{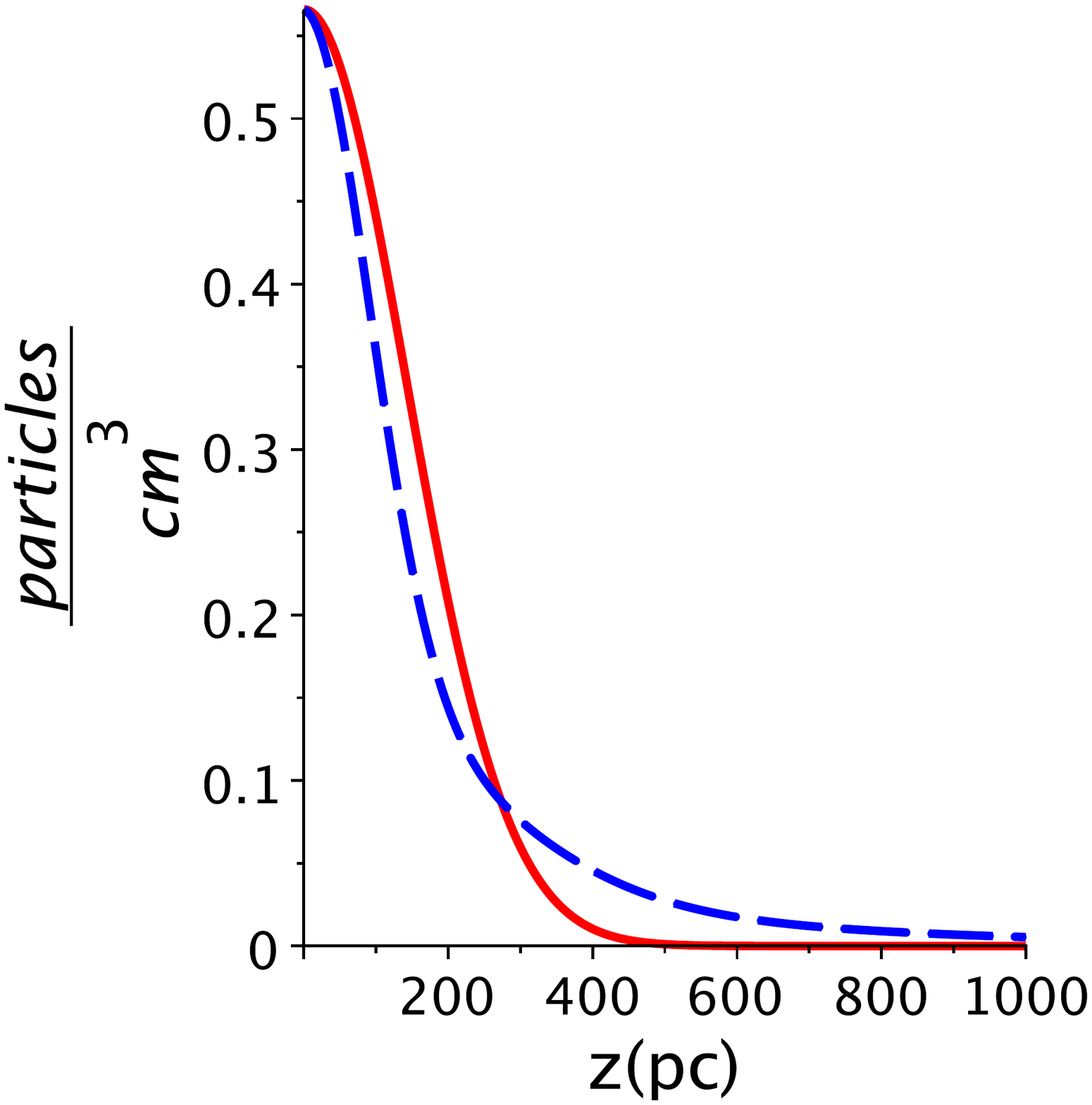}
\end {center}
\caption
{
Profiles of density versus scale height $z$:
the medium is that in the  presence of  a Gaussian medium   as
given by Eq.~(\ref{gaussian})
when $h=90$\ pc
(dashed blue line)
and
the
three-component exponential distribution
as
given by Eq.~(\ref{equation:ism})
(red full line).
}
\label{ism_gauss}
    \end{figure*}

\begin{figure*}
\begin{center}
\includegraphics[width=6cm]{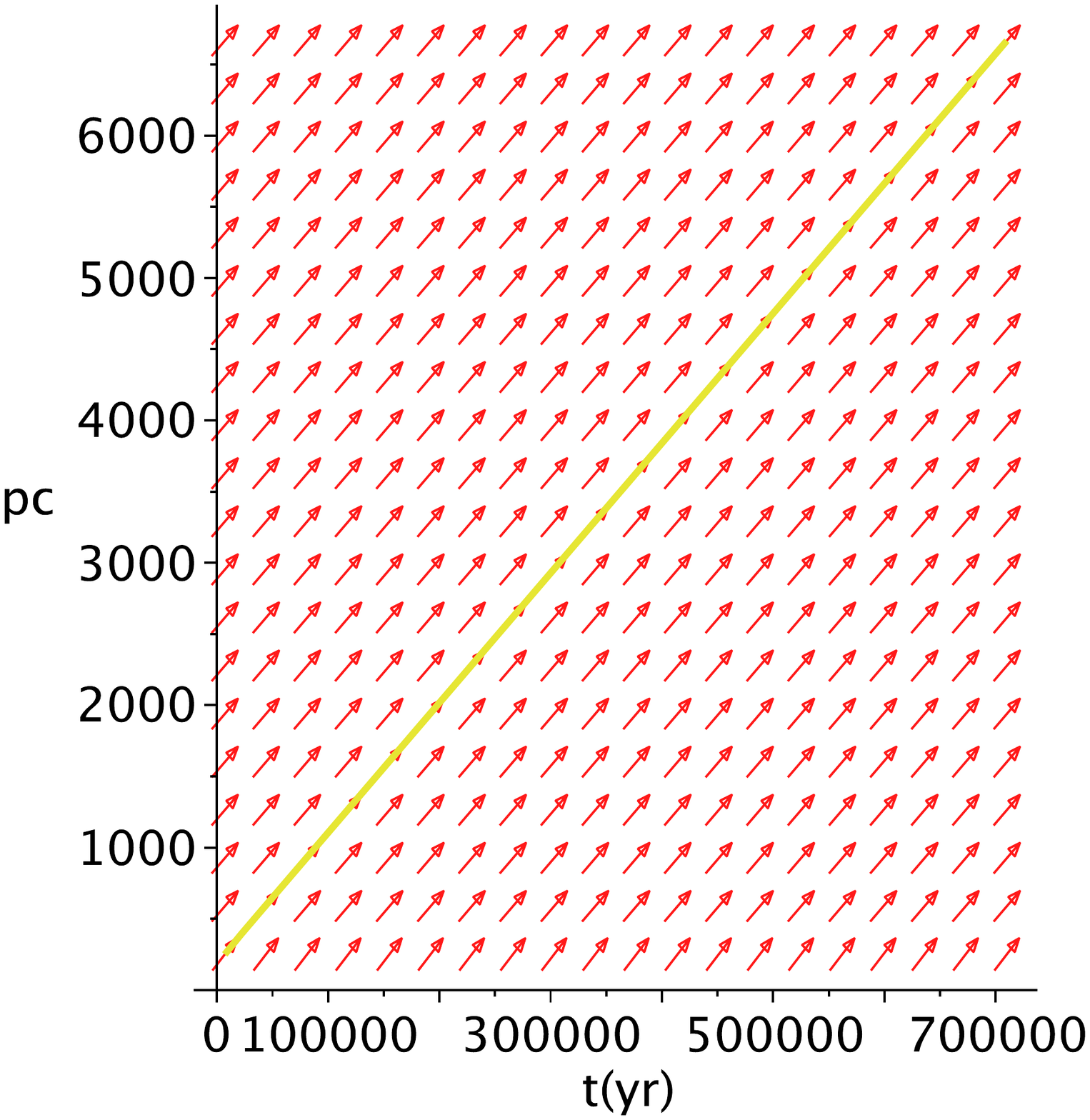}
\end {center}
\caption
{
Phase plane portrait and numerical solution (full yellow line)
in the Gaussian case
when  
$r_0=250$\ pc,
$z_0=90   $\ pc,
$t=7.1\,10^5$\ yr,
$t_0 =7.1 \,10^3$ yr  and
$v_0\,=10000$ km s$^{-1}$.
}
\label{gauss_portrait}
    \end{figure*}

We assume that 
the  number density
distribution scales as   
\begin{equation}
n(z) = n_0 {{\rm e}^{-{\frac {{z}^{2}}{{z_{{0}}}^{2}}}}}
\quad ,
\label{gaussian}
\end{equation}
where $n_0$ is the density at $z=0$
and 
$z_0$ is a scaling parameter.
We now give the adopted  piece-wise dependence for 
the Gaussian medium
\begin{equation}
 \rho (r;r_0,z_0)  =\Bigg \{ \begin{array}{ll}
            \rho_c                      & \mbox {if $r \leq r_0 $ } \\
            \rho_c\,{{\rm e}^{-{\frac {{r}^{2} \left( \cos \left( \theta
\right)  \right) ^{2}}{{{\it z_0}}^{2}}}}}   & \mbox {if $r >     r_0 $}
            \end{array}
\label{piecewisegaussian}
\quad ,
\end{equation}
where  $\rho_c$  is the density at $z=0$.
A  comparison  with 
Eq.~(\ref{equation:ism})
gives  $z_0=200$\ pc,   
see Figure \ref{ism_gauss}. 
The total mass $ M(r;r_0,\rho_c,z_0)$
swept in the interval [0,r]
is
\begin{equation}
M (r;r_0,\rho_c,z_0)= \frac{BN}{12\, \left( \cos \left( \theta \right)
\right) ^{3}}
\quad ,
\end{equation}
where
\begin{eqnarray}
BN=- \Bigg( -4\,{r_{{0}}}^{3} \left( \cos \left( \theta \right)  \right) 
^{3}+3\,{\rm erf} \left({\frac {r_{{0}}\cos \left( \theta \right) }{z_
{{0}}}}\right)\sqrt {\pi}{z_{{0}}}^{3}-3\,{\rm erf} \left({\frac {r
\cos \left( \theta \right) }{z_{{0}}}}\right)\sqrt {\pi}{z_{{0}}}^{3}
\nonumber \\
+
6\,{{\rm e}^{-{\frac {{r}^{2} \left( \cos \left( \theta \right) 
 \right) ^{2}}{{z_{{0}}}^{2}}}}}\cos \left( \theta \right) r{z_{{0}}}^
{2}-6\,{{\rm e}^{-{\frac {{r_{{0}}}^{2} \left( \cos \left( \theta
 \right)  \right) ^{2}}{{z_{{0}}}^{2}}}}}\cos \left( \theta \right) r_
{{0}}{z_{{0}}}^{2} \Bigg ) \rho_{{c}}
\quad ,
\end{eqnarray}
and   erf$(x)$ \cite{NIST2010} is the error function
defined 
by
\begin{equation}
{\rm erf}(x)= 2\,{\frac {\int_{0}^{x}\!{\it \exp} \left( -{t}^{2} \right) \,{\rm
d}
t}{\sqrt {\pi}}}
\quad .
\end{equation}  
The  velocity as a function  of the  radius is
\begin{equation}
v(r;r_0,z_0,v_0) =
\frac{CN}{CD}
\quad ,
\end{equation}
where
\begin{eqnarray}
CN=2\,\Bigg ( {-6\,\cos   ( \theta   )    ( -2/3\,{r_{{0}}}^{3}
   ( \cos   ( \theta   )    ) ^{3}+{{\rm e}^{-{\frac {{r
}^{2}   ( \cos   ( \theta   )    ) ^{2}}{{z_{{0}}}^{2}}}
}}\cos   ( \theta   ) r{z_{{0}}}^{2}-{{\rm e}^{-{\frac {{r_{{0}
}}^{2}   ( \cos   ( \theta   )    ) ^{2}}{{z_{{0}}}^{2}}
}}}\cos   ( \theta   ) r_{{0}}{z_{{0}}}^{2}
}
\nonumber \\
{
-1/2\,{\rm erf} 
  ({\frac {\cos   ( \theta   ) r}{z_{{0}}}}  )\sqrt {\pi
}{z_{{0}}}^{3}+1/2\,{\rm erf}   ({\frac {r_{{0}}\cos   ( \theta
   ) }{z_{{0}}}}  )\sqrt {\pi}{z_{{0}}}^{3}   ) r_{{0}}} \Bigg )^{1/2}
\cos   ( \theta   ) v_{{0}}r_{{0}}
\quad  ,
\end{eqnarray}
and  
\begin{eqnarray}
CD=
4\,{r_{{0}}}^{3} \left( \cos \left( \theta \right)  \right) ^{3}-6\,{
{\rm e}^{-{\frac {{r}^{2} \left( \cos \left( \theta \right)  \right) ^
{2}}{{z_{{0}}}^{2}}}}}\cos \left( \theta \right) r{z_{{0}}}^{2}+6\,{
{\rm e}^{-{\frac {{r_{{0}}}^{2} \left( \cos \left( \theta \right) 
 \right) ^{2}}{{z_{{0}}}^{2}}}}}\cos \left( \theta \right) r_{{0}}{z_{
{0}}}^{2}
\nonumber \\
-3\,{\rm erf} \left({\frac {r_{{0}}\cos \left( \theta
 \right) }{z_{{0}}}}\right)\sqrt {\pi}{z_{{0}}}^{3}+3\,{\rm erf} 
\left({\frac {\cos \left( \theta \right) r}{z_{{0}}}}\right)\sqrt {\pi
}{z_{{0}}}^{3}
\quad  .
\end{eqnarray}
The differential equation which governs
the  motion for the Gaussian  medium
is 
\begin{eqnarray}
4\, \left( \cos \left( \theta \right)  \right) ^{3} \left( {\frac 
{\rm d}{{\rm d}t}}r \left( t \right)  \right) ^{2}{r_{{0}}}^{3}-4\,{r_
{{0}}}^{3}{v_{{0}}}^{2} \left( \cos \left( \theta \right)  \right) ^{3
}-6\,r \left( t \right) \cos \left( \theta \right) {{\rm e}^{-{\frac {
 \left( r \left( t \right)  \right) ^{2} \left( \cos \left( \theta
 \right)  \right) ^{2}}{{z_{{0}}}^{2}}}}} \left( {\frac {\rm d}{
{\rm d}t}}r \left( t \right)  \right) ^{2}{z_{{0}}}^{2}
\nonumber \\
+6\,\cos
 \left( \theta \right) {{\rm e}^{-{\frac {{r_{{0}}}^{2} \left( \cos
 \left( \theta \right)  \right) ^{2}}{{z_{{0}}}^{2}}}}} \left( {\frac 
{\rm d}{{\rm d}t}}r \left( t \right)  \right) ^{2}r_{{0}}{z_{{0}}}^{2}
-3\,\sqrt {\pi}{\rm erf} \left({\frac {r_{{0}}\cos \left( \theta
 \right) }{z_{{0}}}}\right) \left( {\frac {\rm d}{{\rm d}t}}r \left( t
 \right)  \right) ^{2}{z_{{0}}}^{3}
\nonumber \\
+3\,\sqrt {\pi}{\rm erf} \left({
\frac {r \left( t \right) \cos \left( \theta \right) }{z_{{0}}}}
\right) \left( {\frac {\rm d}{{\rm d}t}}r \left( t \right)  \right) ^{
2}{z_{{0}}}^{3}
=0 
\quad .
\label{eqndiffgauss}
\end{eqnarray}
Figure \ref{gauss_portrait}   shows  
the numerical solution obtained with the 
Runge–Kutta method.

\begin{figure*}
\begin{center}
\includegraphics[width=6cm]{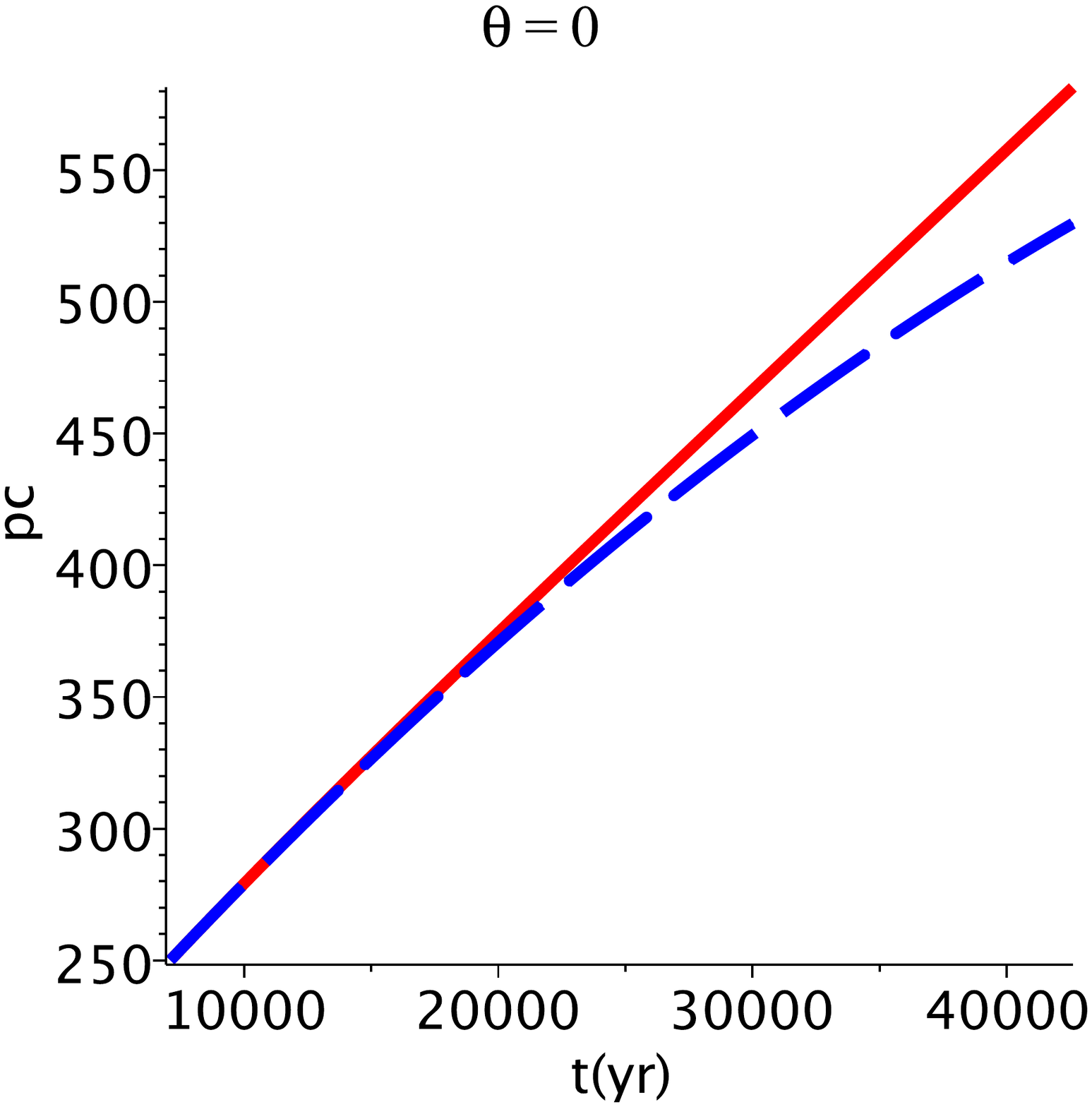}
\end {center}
\caption
{
Numerical solution (red full line)    
and Taylor expansion of the solution (dashed blue line),
parameters as in Figure \ref{gauss_portrait}
but $t=4.26\,10^4$\ yr, 
}
\label{gauss_taylor}
    \end{figure*}

A Taylor expansion of order 3 
of Eq.~(\ref{eqndiffgauss}) 
gives 
\begin{equation}
r(t;t_0,r_0,v_0,z_0)=
r_{{0}}+v_{{0}} \left( t-t_{{0}} \right) -{\frac {3\,{v_{{0}}}^{2}
 \left( t-t_{{0}} \right) ^{2}}{4\,r_{{0}}}{{\rm e}^{-{\frac {{r_{{0}}
}^{2} \left( \cos \left( \theta \right)  \right) ^{2}}{{z_{{0}}}^{2}}}
}}}
\quad ,
\end{equation}
and  Figure 
\ref{gauss_taylor}   
gives  
the numerical solution obtained by the 
Runge–Kutta method
and  the series solution
up to a time for which the percentage error  is 
less than $9\%$.

\section{Astrophysical  applications}

\label{section_astrophysical}

In the  following we will    analyse  
the  local bubble, the Fermi bubble and
the super bubble W4.
An observational
percentage reliability, $\epsilon_{\mathrm {obs}}$,
is  introduced over the whole range
of the polar   angle  $\theta$,
\begin{equation}
\epsilon_{\mathrm {obs}}  =100(1-\frac{\sum_j |r_{\mathrm {obs}}-r_{\mathrm{num}}|_j}{\sum_j
{r_{\mathrm {obs}}}_{,j}})
\quad,
\label{efficiencymany}
\end{equation}
where
$r_{\mathrm{num}}$ is the theoretical radius of the considered bubble,
$r_{\mathrm{obs}}$ is the observed    radius of the considered bubble, 
and
the  index $j$  varies  from 1 to the number of
available observations.
The observational
percentage of reliability  allows us to fix the theoretical parameters.

\subsection{The local bubble}

\begin{figure*}
\begin{center}
\includegraphics[width=5cm]{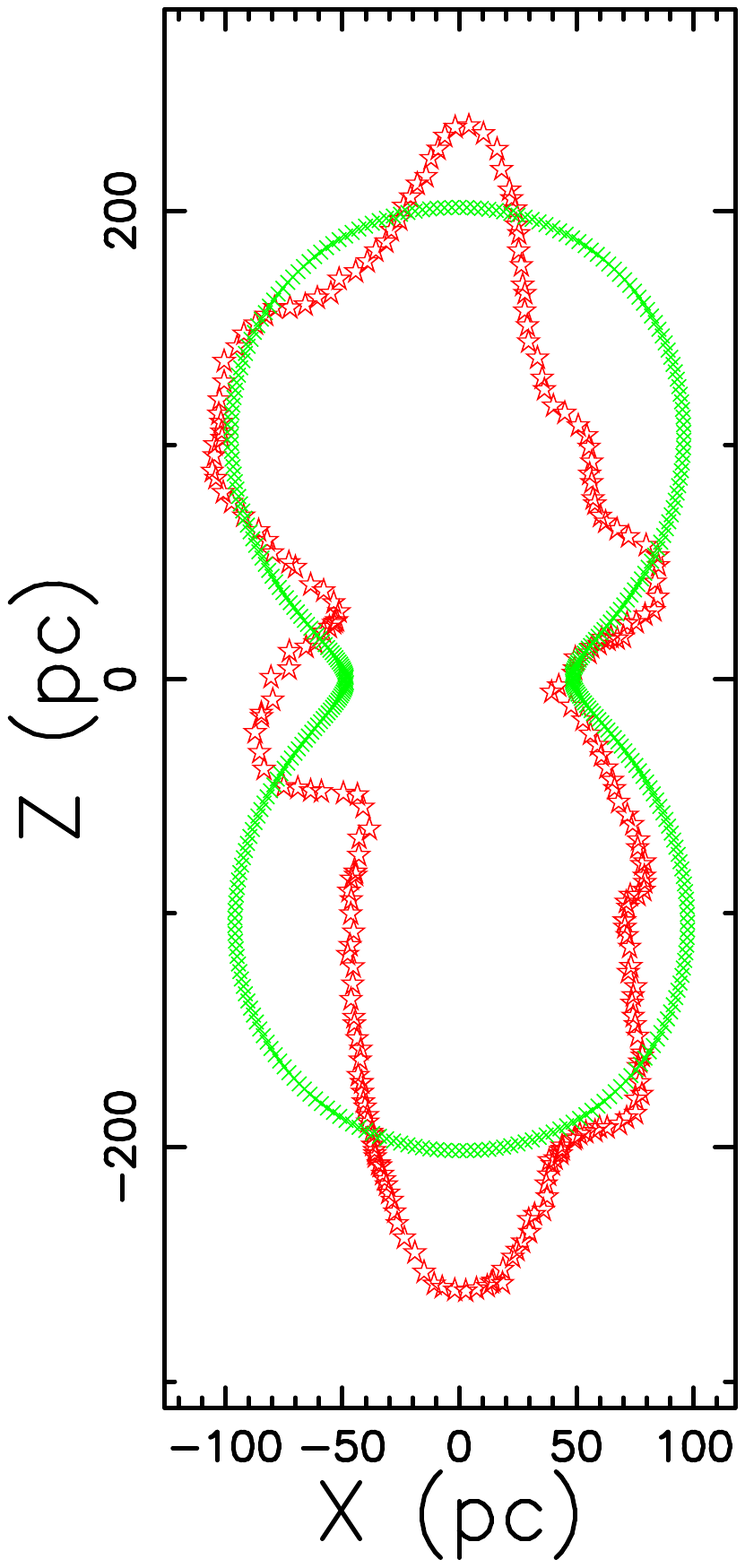}
\end {center}
\caption
{
Geometrical section of the LB 
in the $x-z$ plane with a profile
in  the presence of 
self-gravity
as
given by Eq.~(\ref{sech2})
(green points)
and observed profile
(red stars).
The parameters are 
$v_0\,=3700$ km s$^{-1}$,
$r_0\,=7$\ pc,
$h   \,=3.5$\ pc,
$t=8.5\,10^4$\ yr, 
$t_0=8.5\,10^2$\ yr
and $z_{OB}=0$. 
The observational
percentage reliability is 
$\epsilon_{\mathrm {obs}}=82.42\%$.
}
\label{localb_theo_obs_sech2}
    \end{figure*}

\begin{figure*}
\begin{center}
\includegraphics[width=5cm]{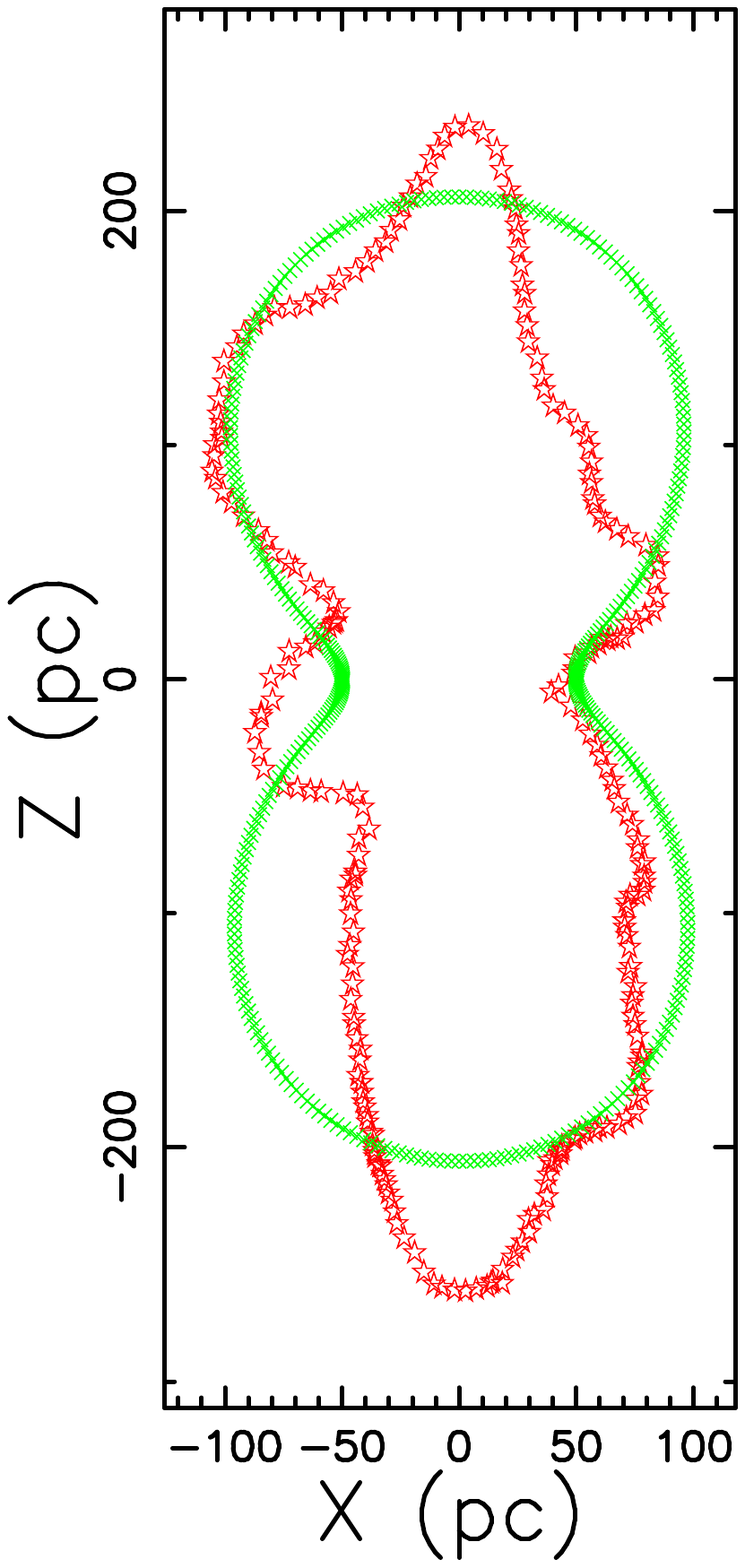}
\end {center}
\caption
{
Geometrical section of the LB 
in the $x-z$ plane with a Gaussian  profile
(green points)
and observed profile
(red stars).
The parameters are 
$v_0\,=4000$ km s$^{-1}$,
$r_0\,=7 $\ pc,
$z_0 \,=9.1 $\ pc,
$t=8.5\,10^4$\ yr,
$t_0=8.5\,10^2 $\ yr
and $z_{OB}=0$ .
The observational
percentage reliability is $\epsilon_{\mathrm {obs}}=82.58\%$.
}
\label{localb_theo_obs_gauss}
    \end{figure*}

The local bubble (LB) has already been simulated 
in the framework of the conservation of momentum 
\cite{Zaninetti2020e};  here  we 
adopt the framework of the conservation of energy.
The  numerical solution is shown 
as a cut   in 
the $x-z$ plane:
see Figure \ref{localb_theo_obs_sech2} 
for a  medium in  the presence of 
self-gravity   
as given by Eq.~(\ref{piecewisesech2})
and
Figure \ref{localb_theo_obs_gauss} 
for a Gaussian  density profile
as given by Eq.~(\ref{piecewisegaussian}).

\begin{figure*}
\begin{center}
\includegraphics[width=5cm]{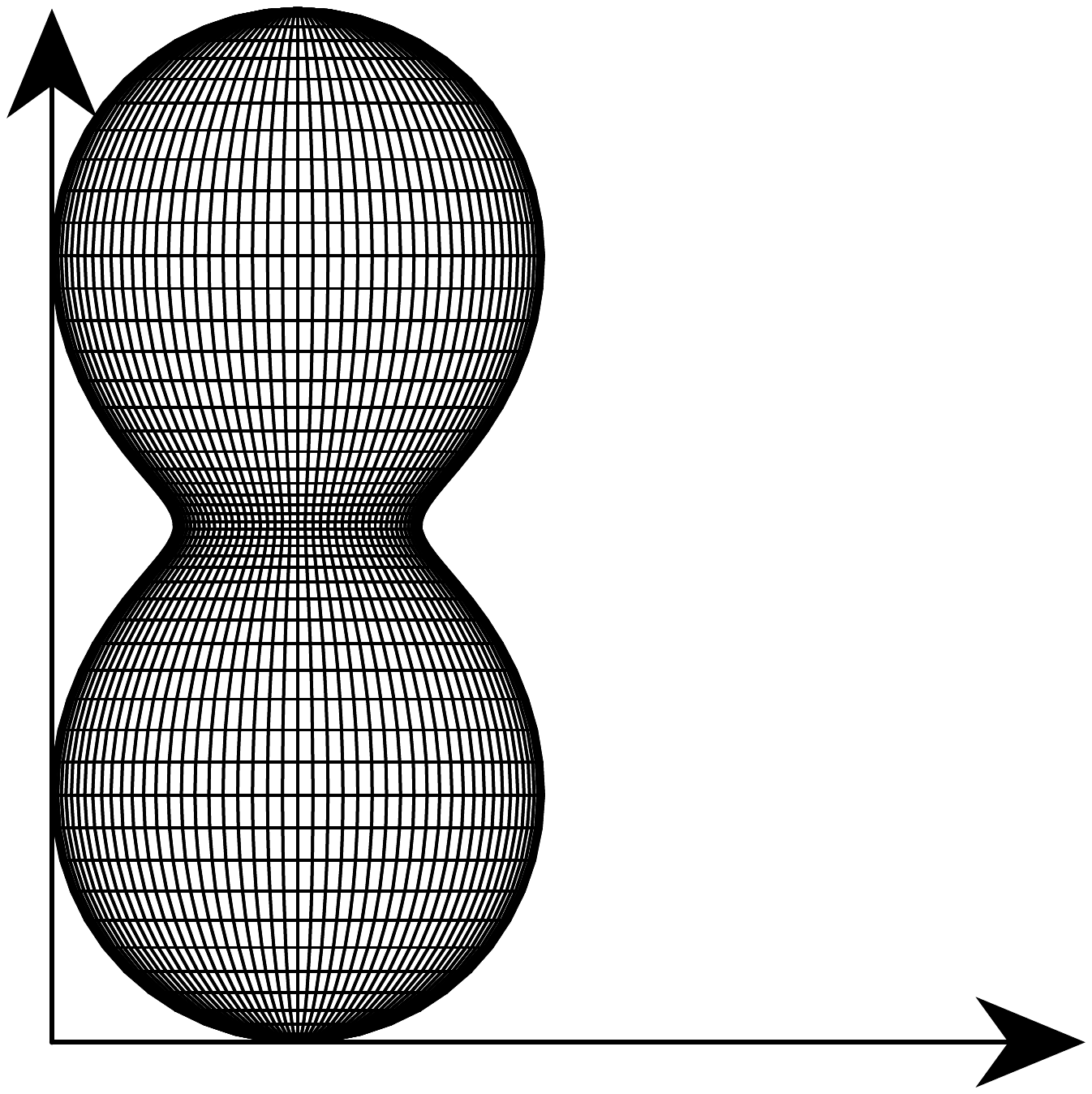}
\end {center}
\caption
{
3D surface  of the LB 
with parameters as in Figure  \ref{localb_theo_obs_sech2}, 
with a  profile  in  presence of 
self-gravity
as
given by Eq.~(\ref{sech2}).
The three Euler angles are $\Theta=90^{\circ}$, $\Phi=0^{\circ}$ and
$ \Psi=90^{\circ}$.
}
\label{localb_sech2_3d_bn}
    \end{figure*}

The 3D advancing surface  of the local bubble 
for the case of 
self-gravity
is shown 
in Figure \ref{localb_sech2_3d_bn}.

\subsection{The Fermi  bubble}

\begin{figure*}
\begin{center}
\includegraphics[width=5cm]{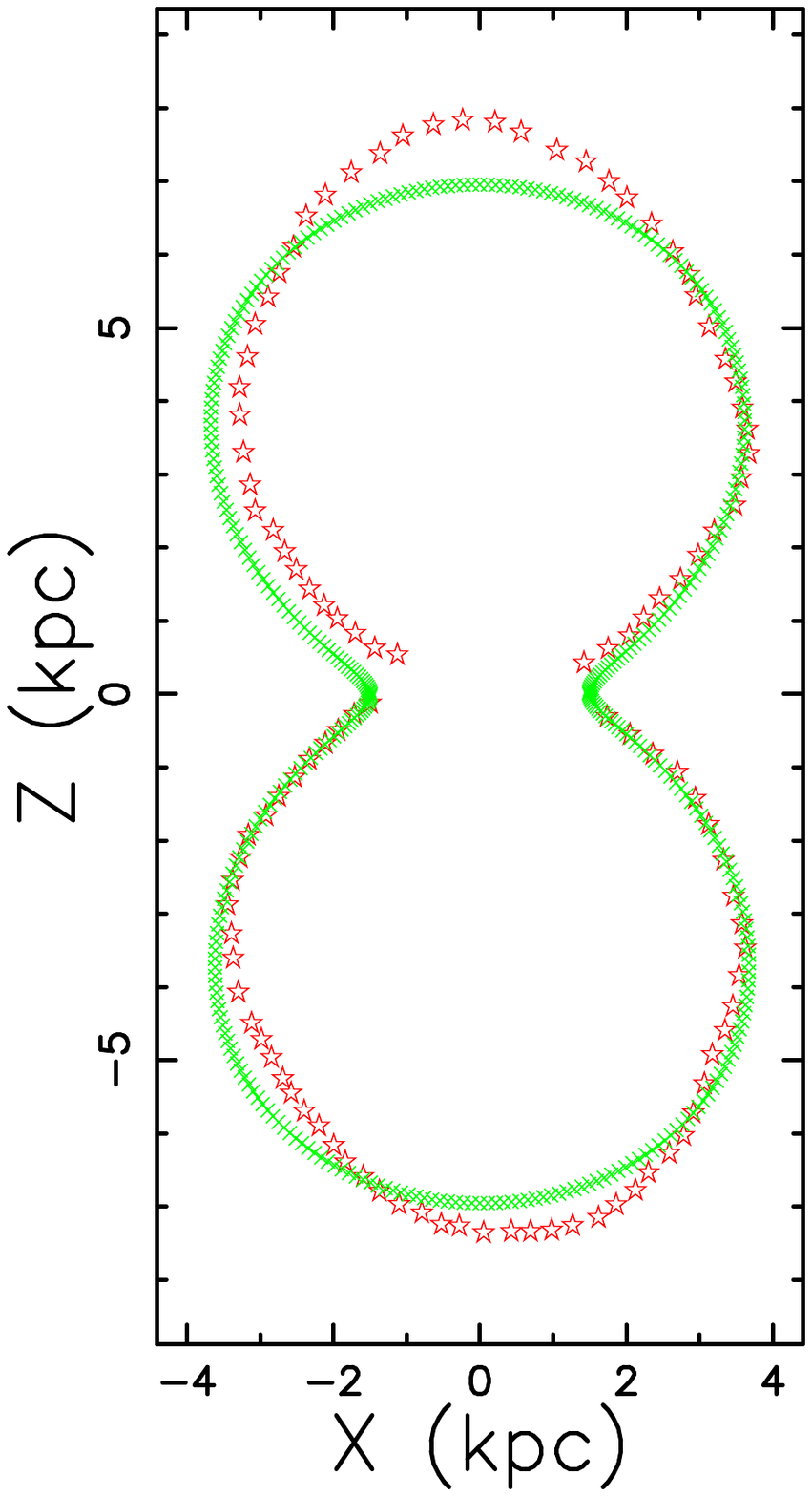}
\end {center}
\caption
{
Geometrical section of the Fermi bubbles  
in the $x-z$ plane with a profile
in  the presence of 
self-gravity
as
given by Eq.~(\ref{sech2})
(green points)
and observed profile
(red stars).
The parameters are 
$v_0\,=2000$ km s$^{-1}$,
$r_0\,=250 $\ pc,
$z_0\,=12  $\ pc,
$t=4 \,10^6$\ yr,
$t_0=4 \,10^4 $\ yr
and $z_{OB}=0$.
The observational
percentage reliability is $\epsilon_{\mathrm {obs}}=93\%$.
}
\label{fermisb_theo_obs_sech2}
    \end{figure*}

\begin{figure*}
\begin{center}
\includegraphics[width=5cm]{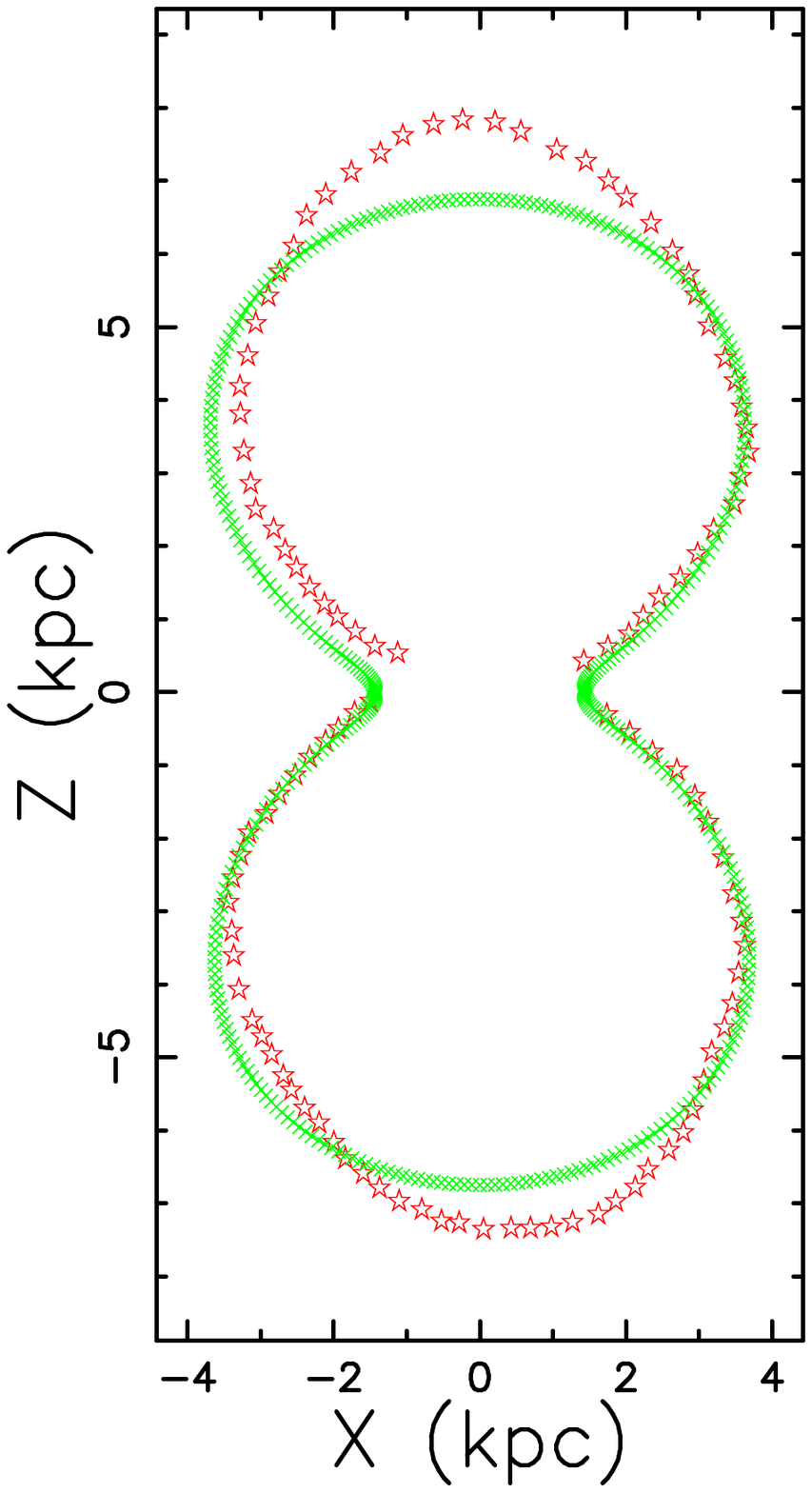}
\end {center}
\caption
{
Geometrical section of the  Fermi bubbles
in the $x-z$ plane with a Gaussian  profile
(green points)
and observed profile
(red stars).
The parameters are 
$v_0\,=1000$ km s$^{-1}$,
$r_0\,=250 $\ pc,
$z_0 \,=200 $\ pc,
$t=7.1\,10^5$\ yr,
$t_0=7.1\,10^3 $\ yr
and $z_{OB}=0$.
The observational
percentage reliability is $\epsilon_{\mathrm {obs}}=92.12\%$.
}
\label{fermisb_theo_obs_gauss}
    \end{figure*}

Fermi bubbles have already been simulated
in the framework of the conservation  of momentum 
\cite{Zaninetti2018c}; 
here we apply the conservation of energy.
We now test  our models 
on  the image of the Fermi bubbles 
available at
\url{https://www.nasa.gov/mission_pages/GLAST/news/new-structure.html}.
The  numerical solution is shown 
as a cut   in 
the $x-z$ plane:
see Figure \ref{fermisb_theo_obs_sech2} 
for a density profile
in  the presence of 
self-gravity

\noindent as given by Eq.~(\ref{piecewisesech2})
and
Figure \ref{fermisb_theo_obs_gauss} 
for a Gaussian  density profile
as given by Eq.~(\ref{piecewisegaussian}).

\begin{figure*}
\begin{center}
\includegraphics[width=5cm,angle=-90]{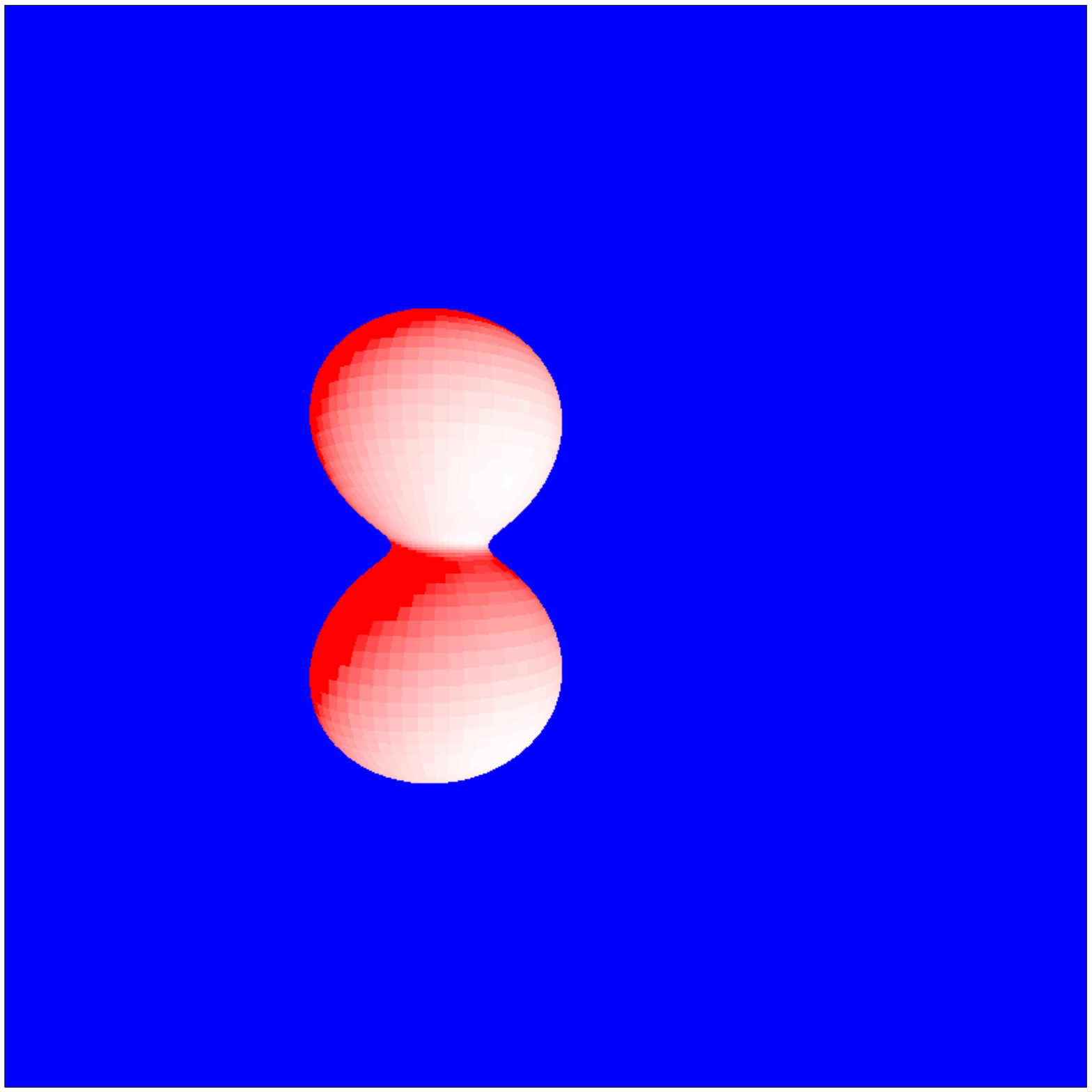}
\end {center}
\caption
{
3D surface  of the Fermi bubbles 
with parameters as in Figure  \ref{fermisb_theo_obs_gauss}, 
Gaussian profile.
The three Euler angles are $\Theta=90^{\circ}$, $\Phi=0^{\circ}$ and
$ \Psi=90^{\circ}$.
}
\label{fermisb_gauss_3d}
    \end{figure*}

The 3D advancing surface  of the local bubble 
for the Gaussian  case  is shown 
in Figure \ref{fermisb_gauss_3d}.

\subsection{The W4 super-bubble}

\begin{figure*}
\begin{center}
\includegraphics[width=5cm]{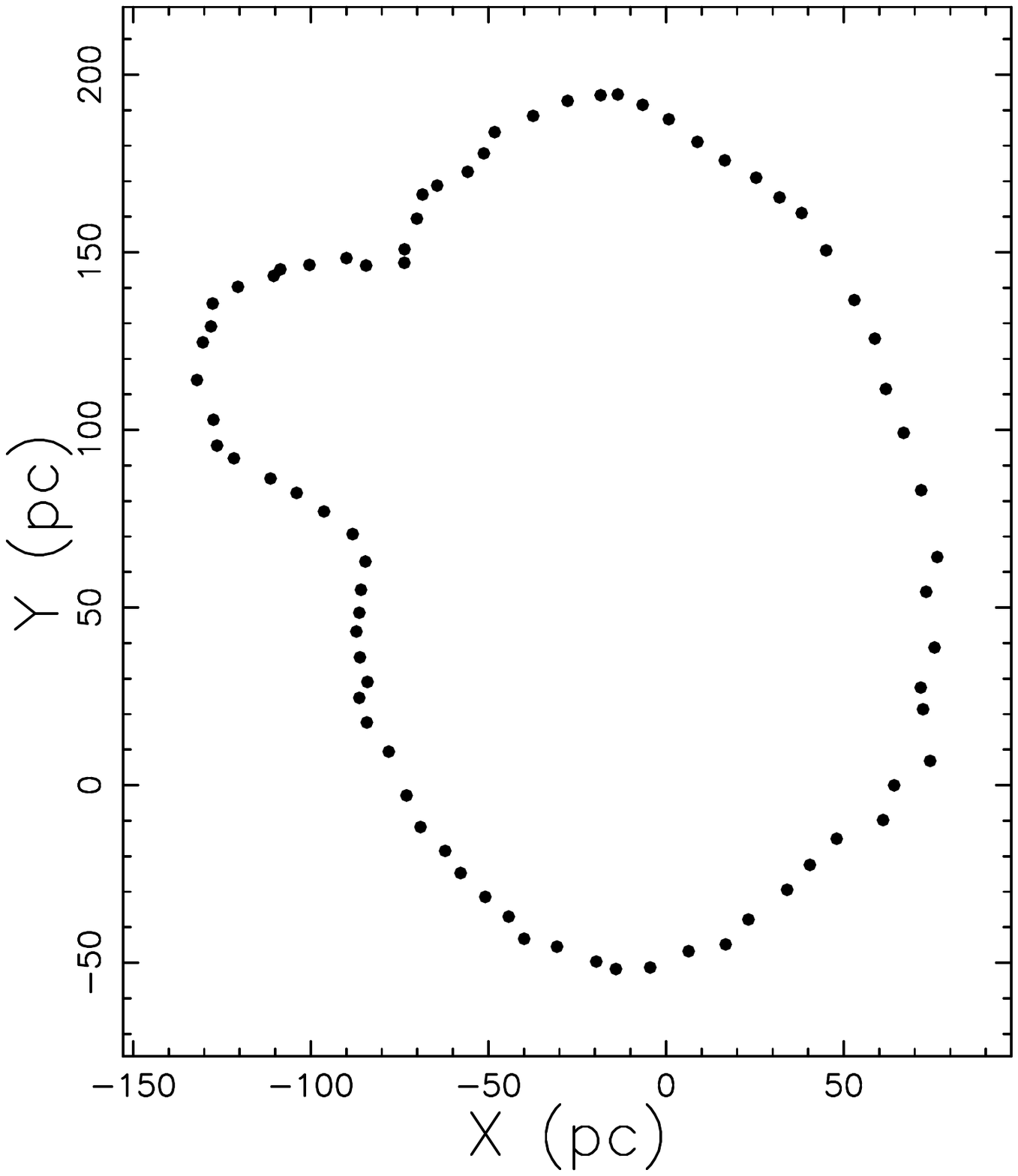}
\end {center}
\caption
{
Section of the W4 + IC 1805 complex.
}
\label{w4_observed}
    \end{figure*}

The  W4  super-bubble has been 
analysed  from  the point
of view of the astronomical observations 
\cite{Normandeau1999,Normandeau2000,West2007},
in connection with the evolution of the magnetic field
\cite{Gao2015} and from   a theoretical point of view
\cite{Basu1999,Baumgartner2013}. 
The upper part of Figure 3 in \cite{Megeath2008}, which combines 
[SII], $H\alpha$ and  [OIII] images has  been digitized
and will be  the section  of reference for W4,
see  Figure \ref{w4_observed}.

\begin{figure*}
\begin{center}
\includegraphics[width=5cm]{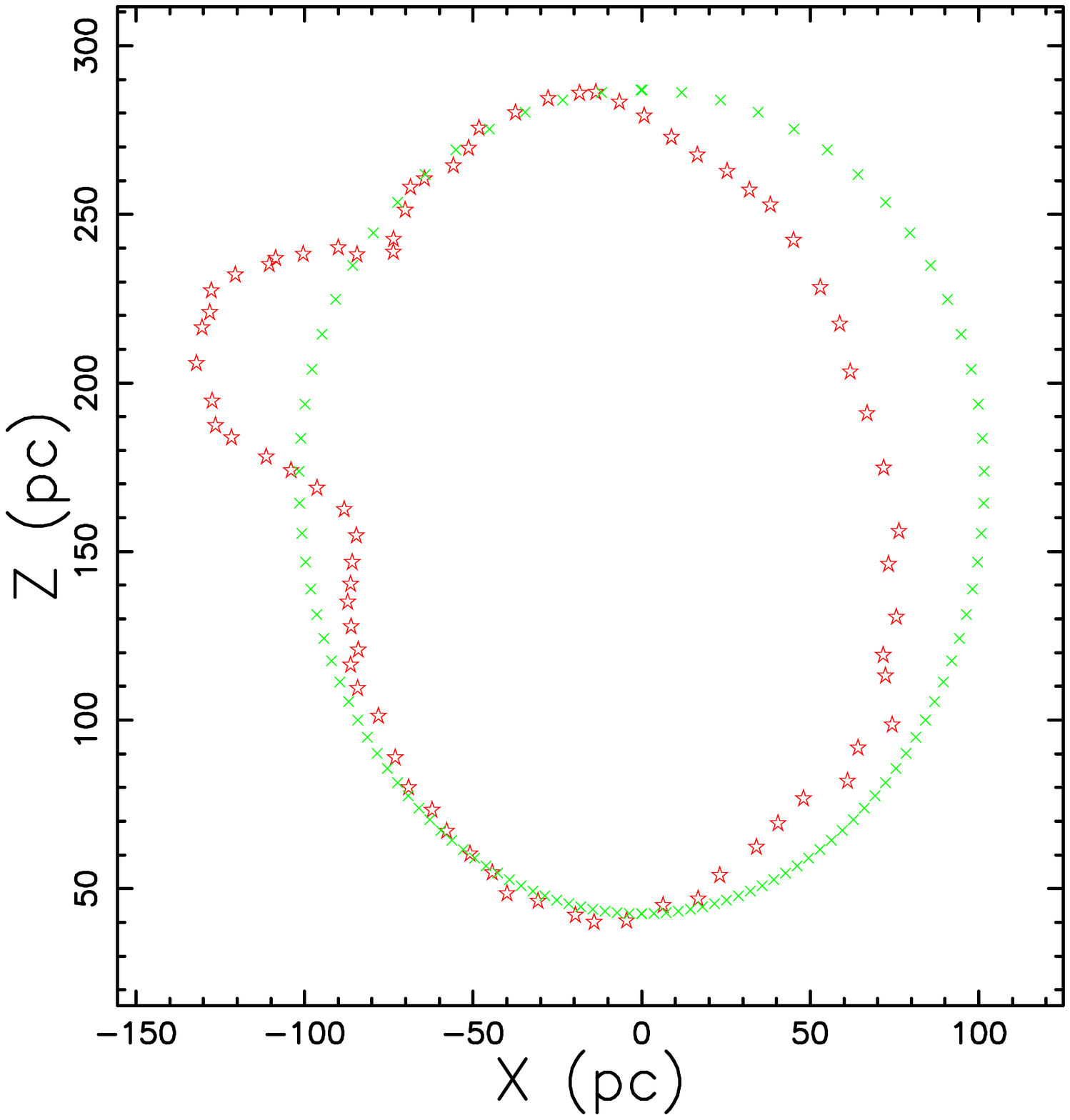}
\end {center}
\caption
{
Geometrical section of the W4 super-bubble  
in the $X-Z$ plane with a
profile
in  the presence of 
self-gravity
as
given by Eq.~(\ref{sech2})
(green points)
and observed profile
(red stars).
The parameters are 
$v_0\,=20000$ km s$^{-1}$,
$r_0\,=1 $\ pc,
$h \,=17  $\ pc,
$t=1.3 \,10^5$\ yr,
$t_0=10 $\ yr
and $z_{OB}=100$.
}
\label{theo_obs_w4_sech2}
    \end{figure*}

\begin{figure*}
\begin{center}
\includegraphics[width=5cm]{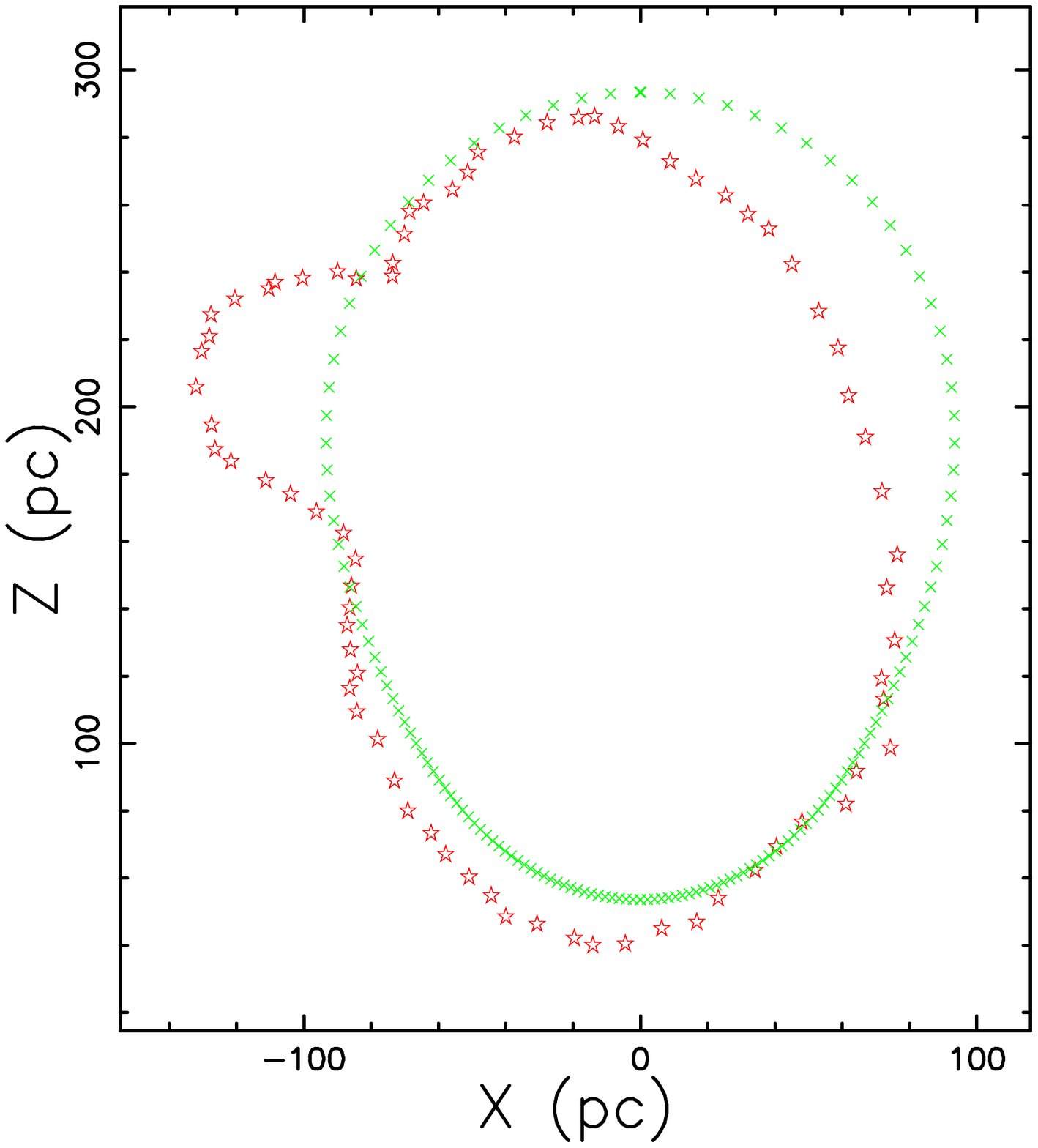}
\end {center}
\caption
{
Geometrical section of the W4 super-bubble  
in the $X-Z$ plane with a  Gaussian   profile
(green points)
and observed profile
(red stars).
The parameters are 
$v_0\,=4700$ km s$^{-1}$,
$r_0\,=1 $\ pc,
$z_0 \,=55  $\ pc,
$t=2.25 \,10^5$\ yr,
$t_0=10 $\ yr
and $z_{OB}=100$.
}
\label{theo_obs_w4_gauss}
    \end{figure*}

We now  simulate the egg-shape of W4
when $z_{OB} \neq 0$. 
The  numerical solution, which is
evaluated with the Euler method,
 is shown 
as a cut   in 
the $x-z$ plane:
see Figure \ref{theo_obs_w4_sech2} 
for a density profile
in the  presence of 
self-gravity
and  Figure \ref{theo_obs_w4_gauss} 
for a Gaussian profile.
The two adopted  profiles in density 
are  symmetric with respect to the galactic plane, $Z=0$,  
but  the simulated  theoretical sections do not have an up--down symmetry, due to the fact
that the expansion starts at  $z=z_0$.
Nevertheless, we still have a right--left symmetry.   

\begin{figure*}
\begin{center}
\includegraphics[width=5cm,angle=+90]{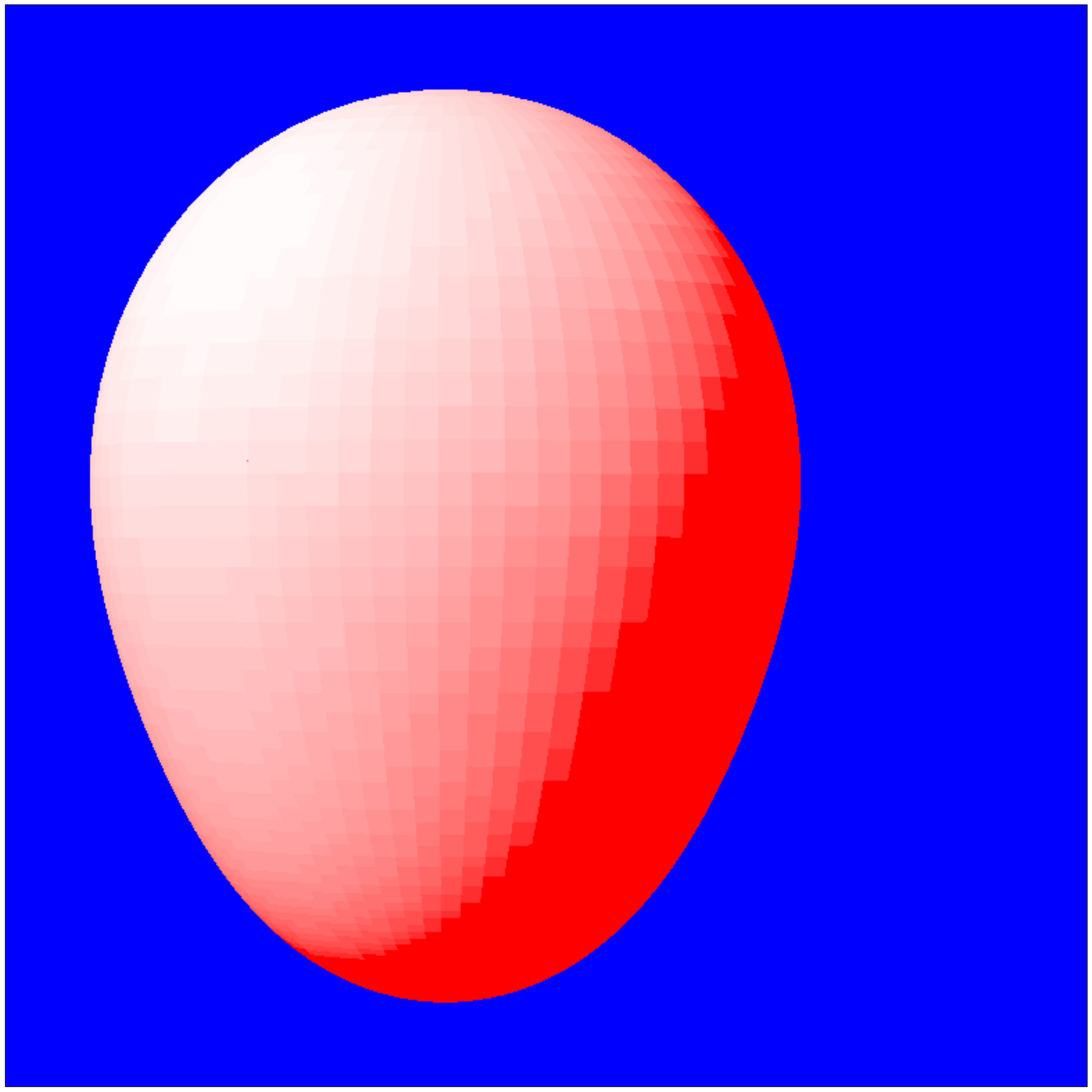}
\end {center}
\caption
{
3D surface  of the W4 super-bubble  
with parameters as in Figure  \ref{theo_obs_w4_gauss}, 
Gaussian profile.
The three Euler angles are $\Theta=90^{\circ}$, $\Phi=0^{\circ}$ and
$ \Psi=90^{\circ}$.
}
\label{w4_3d_gauss}
    \end{figure*}

The egg shape  of the W4 super-bubble  is shown 
in Figure \ref{w4_3d_gauss}.

The curious bump  visible in the upper left part  of
Figure \ref{w4_observed} could be   an astronomical  superposition
of the image of IC 1805  on W4  or 
an intrinsic feature in  the expansion of W4.
In order to  reproduce  this feature, we assume 
that  the   scaling  factor $z_{0,\theta}$
in the interval $\theta_{inf} < \theta < \theta_{sup}$ 
varies 
with the following empirical law  
\begin{equation} 
z_{0,\theta} =z_0 + z0 \, 0.0006 N(\theta;\overline{\theta},\sigma) 
\end{equation}
where   
\begin{equation}
N(\theta;\sigma,\mu ) =
\frac {1} {\sigma (2 \pi)^{1/2}} \exp ({- {\frac {(x-\overline{\theta})^2}{2\sigma^2}}} )
\quad ,
\label{gaussiann}
\end{equation}
is the Gaussian distribution,
and  
$\overline{\theta} = \frac{\theta_{inf} +\theta_{sup}}{2}$ 
and $\sigma=\overline{\theta}/9$.

\begin{figure*}
\begin{center}
\includegraphics[width=5cm]{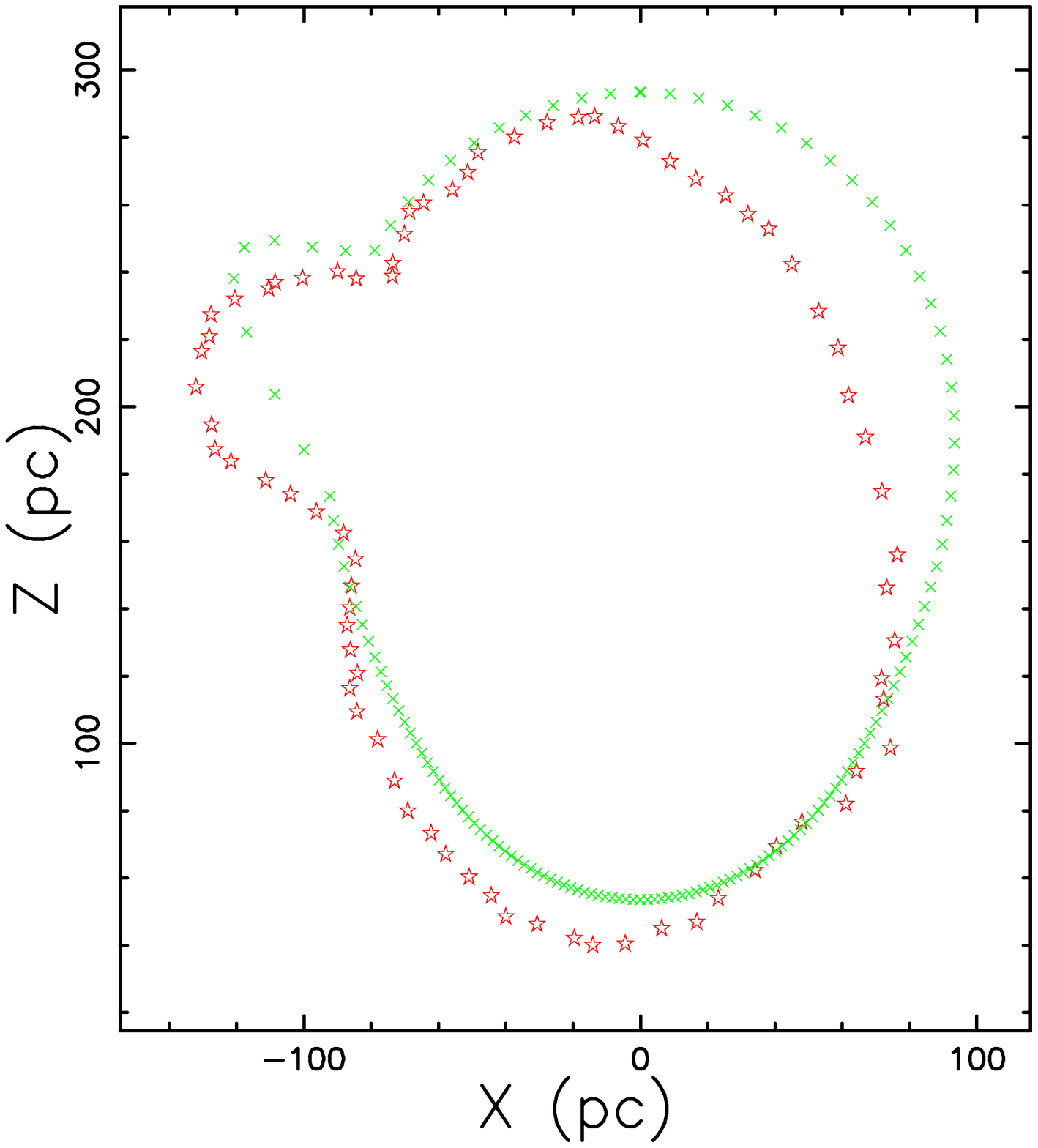}
\end {center}
\caption
{
Geometrical section of the W4 super-bubble  
in the $X-Z$ plane with a  Gaussian   profile
(green points)
and observed profile
(red stars).
The parameters are the same as for Figure 
\ref{theo_obs_w4_gauss}.
}
\label{theo_bump_w4}
    \end{figure*}

Figure \ref{theo_bump_w4} shows  an 
`ad hoc' simulation of the  bump of  W4.

\section{The theory of the image}
\label{section_image}

In the framework of an optically 
thin medium, 
we outline 
a new analytical  model 
which reproduces  a theoretical vertical cut
in the intensity of radiation 
and 
an old numerical model which  
simulates the intensity of radiation 
as a function 
of the point of view of the observer.
    
\subsection{The piriform model}

\begin{figure*}
\begin{center}
\includegraphics[width=6cm]{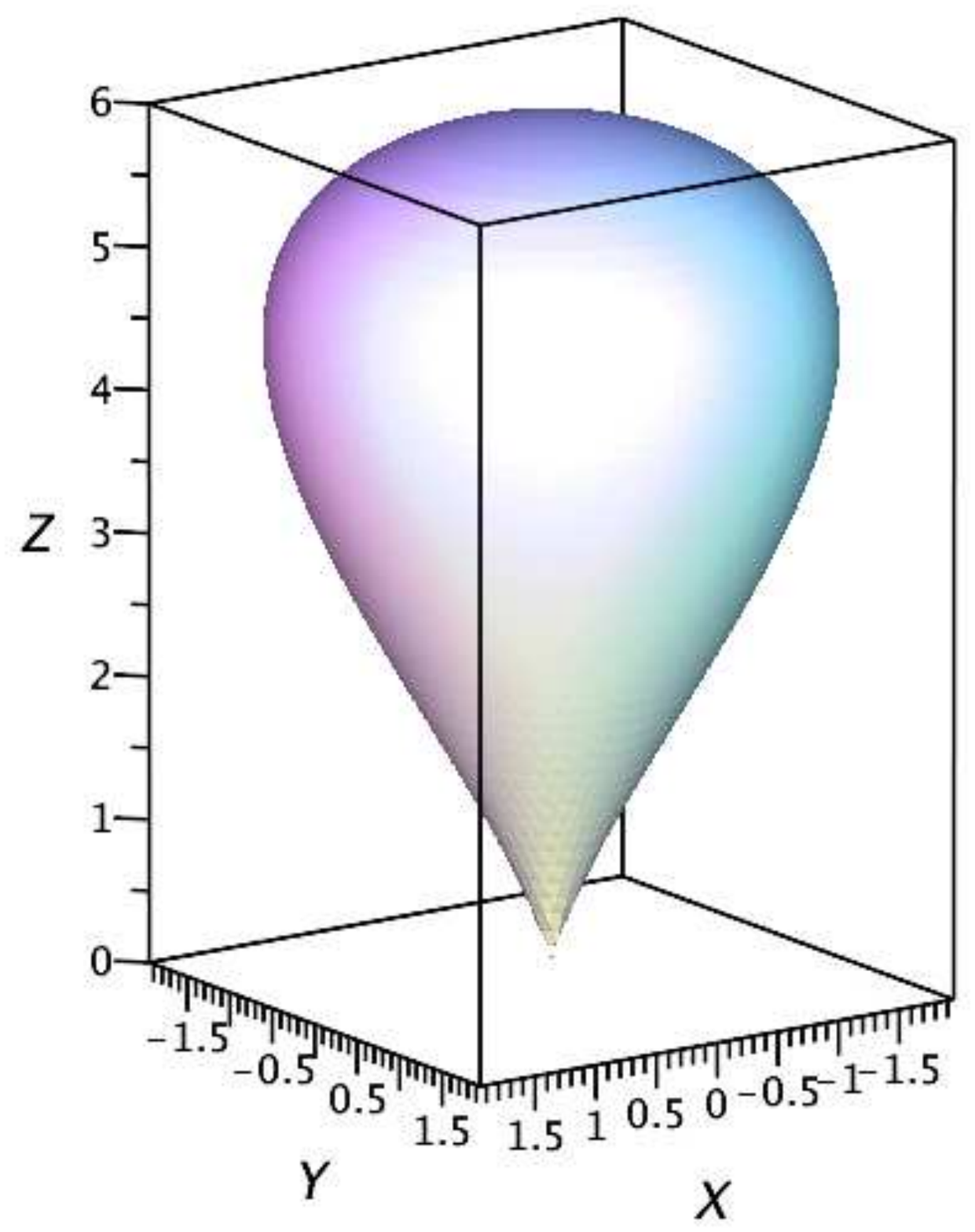}
\end {center}
\caption
{
3D display   of the piriform curve 
when $a=3$\ kpc and $b=3/2$\ kpc.
}
\label{piriform3d}
    \end{figure*}

The  piriform curve, or pear-shaped quartic, in 3D 
Cartesian coordinates $(x,y,z)$ 
has  the equation 
\begin{equation}
{a}^{4} \left( {x}^{2}+{y}^{2} \right) -{b}^{2}{z}^{3} \left( 2\,a-z
 \right) 
=0 
\quad ,
\end{equation}
where $a$ and $b$ are both positive \cite{Lawrence2013},
see Figure  \ref{piriform3d} 
where the parameters $a$ and $b$ match 
the  Fermi bubbles.

\begin{figure*}
\begin{center}
\includegraphics[width=7cm]{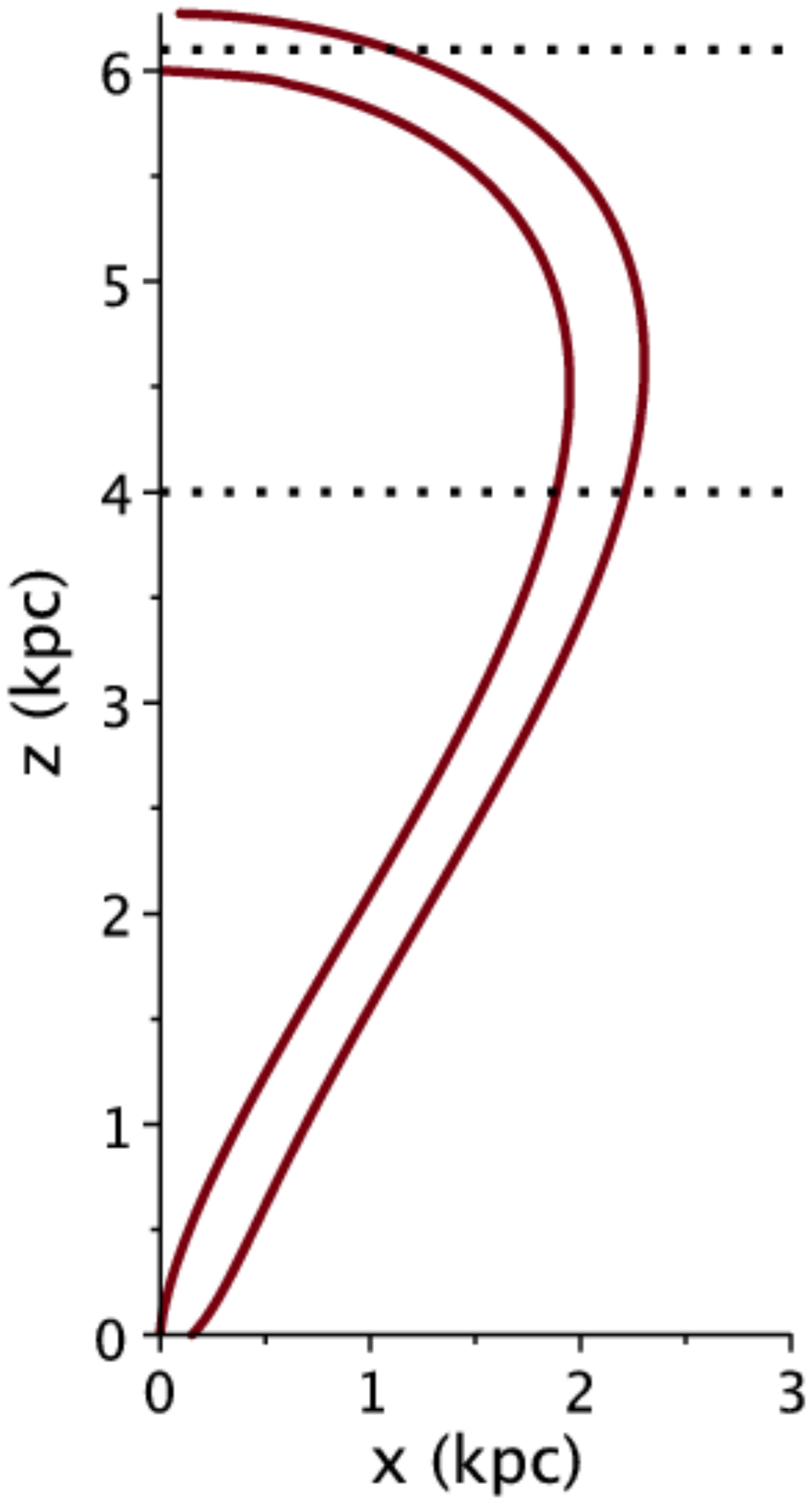}
\end {center}
\caption
{
Internal and external piriforms in the first 
quadrant  
when $a=3$\ kpc,
$b=3/2$\ kpc
and 
$c=3/20$.
The dotted lines represent two different 
lengths of sight.
}
\label{twopiriform}
    \end{figure*}

\begin{figure*}
\begin{center}
\includegraphics[width=7cm]{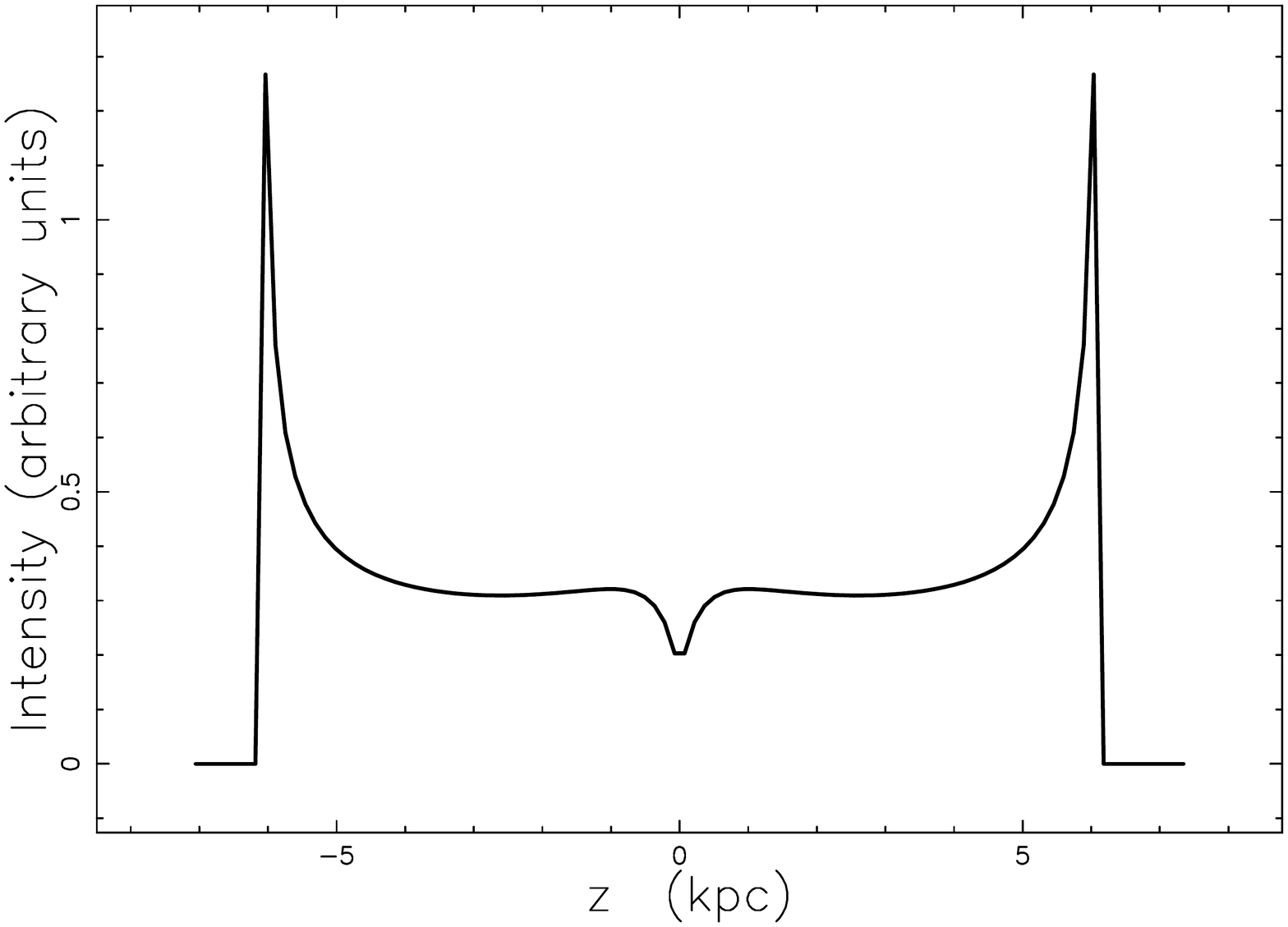}
\end {center}
\caption
{
The intensity profile for the  piriform model 
along the $z$-axis when  
when $a=3$\ kpc,
$b=3/2$\ kpc,
$c=3/20$
and
$C_m$=1.
}
\label{cut_piriform}
    \end{figure*}

We are interested in a section of the above curve  
in the $x-z$ plane 
which is obtained 
by inserting $y=0$
\begin{equation}
{a}^{4}{x}^{2}={b}^{2}{z}^{3} \left( 2\,a-z \right)
\quad .
\end{equation}
The parametric form of the piriform curve is
\begin{subequations}
\begin{align}
x(\theta;a,b) =a \left( 1+\sin \left( \theta \right)  \right) 
\\
z(\theta;a,b)=b\cos \left( \theta \right)  \left( 1+\sin \left( \theta
\right) 
 \right) 
\quad ,
\end{align}
\end{subequations}
where  $-\frac{\pi}{2} \leq \theta \leq \frac{3 \pi}{2}$ 
and 
the maximum value reached  
along the $z$ axis is  
\begin{equation}
z_{max}=2\,a
\quad .
\end{equation}
We assume 
that the emission takes place in a  thin layer comprised between
an internal  piriform
which in polar coordinates has radius    
\begin{equation} 
r_{int} ={\frac {z}{{a}^{2}}\sqrt {{a}^{4}+2\,a{b}^{2}z-{b}^{2}{z}^{2}}}
\quad ,
\label{piriform_int}
\end{equation}
and an external piriform
which has radius
\begin{equation} 
r_{ext} =r_{int} +c  
\quad ,
\label{piriform_ext}
\end{equation}
where $c$ is a positive parameter,
see Figure \ref{twopiriform}.
We therefore
assume that the number density $C_m$ is constant between the
two piriforms;
as  an example, along the $z$ axis the number density    
increases from 0 at  $(0,z_{max})$  to a maximum value $C_m$, 
remains constant
up to $(0,z_{max}+c)$, and then falls again to 0. 
The length of sight which produces the image in the first quadrant, 
when
the observer is situated at the infinity of the $x$-axis, is the
locus parallel to the $x$-axis which  crosses  the position $z$ in
the Cartesian $x-z$ plane and terminates at the external 
piriform.
In the case of an optically thin medium, 
the  line of sight   is split into two cases  
\begin{eqnarray}
l_I(z;a,b,c,C_m)  =
\nonumber  \\
\frac{1}{{a}^{2}} \Bigg ( 
C_{{m}}\sqrt {{a}^{4}{c}^{2}+2\,{b}^{2}{z}^{3}a-{b}^{2}{z}^{4}+2\,
\sqrt { \left( 2\,az-{z}^{2} \right) {b}^{2}+{a}^{4}}{a}^{2}cz}
\Bigg )  
\\
when \quad   z_{max} \leq z < z_{max} +c  \nonumber  \\
l_{II}(z;a,b,c,C_m) =
\nonumber\\ 
\frac{1}{{a}^{2}} \Bigg (  
-C_{{m}} \left( \sqrt {z \left( 2\,a-z \right) }zb-\sqrt {{a}^{4}{c}^{
2}+2\,{b}^{2}{z}^{3}a-{b}^{2}{z}^{4}+2\,\sqrt { \left( 2\,az-{z}^{2}
 \right) {b}^{2}+{a}^{4}}{a}^{2}cz} \right) 
\Bigg )
\\
when \quad  0  \leq z < z_{max}  
\nonumber  
\label{linesigth}
\quad  .
\end{eqnarray}
A comparison between observed and  theoretical
intensity  can be made  
by replacing  in the above result $C_m$ with $I_m$ 
and doubling the length  of sight 
due to the contribution of the second quadrant
\begin{eqnarray}
I_I(z;a,b,c,I_m)  =
\nonumber  \\
2 \times \frac{1}{{a}^{2}} \Bigg ( 
I_{{m}}\sqrt {{a}^{4}{c}^{2}+2\,{b}^{2}{z}^{3}a-{b}^{2}{z}^{4}+2\,
\sqrt { \left( 2\,az-{z}^{2} \right) {b}^{2}+{a}^{4}}{a}^{2}cz}
\Bigg )  
\\
when \quad   z_{max} \leq z < z_{max} +c  \nonumber  \\
I_{II}(z;a,b,c,I_m) =
\nonumber\\ 
2 \times \frac{1}{{a}^{2}} \Bigg (  
-I_{{m}} \left( \sqrt {z \left( 2\,a-z \right) }zb-\sqrt {{a}^{4}{c}^{
2}+2\,{b}^{2}{z}^{3}a-{b}^{2}{z}^{4}+2\,\sqrt { \left( 2\,az-{z}^{2}
 \right) {b}^{2}+{a}^{4}}{a}^{2}cz} \right) 
\Bigg )
\\
when \quad  0  \leq z < z_{max}  
\nonumber  
\label{lineintensity}
\quad  .
\end{eqnarray}
The   resulting intensity is $I_{{m}}\,2\,c$  at $z=0$
and increases to $I_{{m}}\,2\,\sqrt {c}\sqrt {4\,a+c}$ 
at  $z=z_{max}$,
see  Figure \ref{cut_piriform}
for a  typical profile in intensity along the $z$-axis.

\subsection{The numerical   model}

\begin{figure}
\includegraphics[width=6cm]{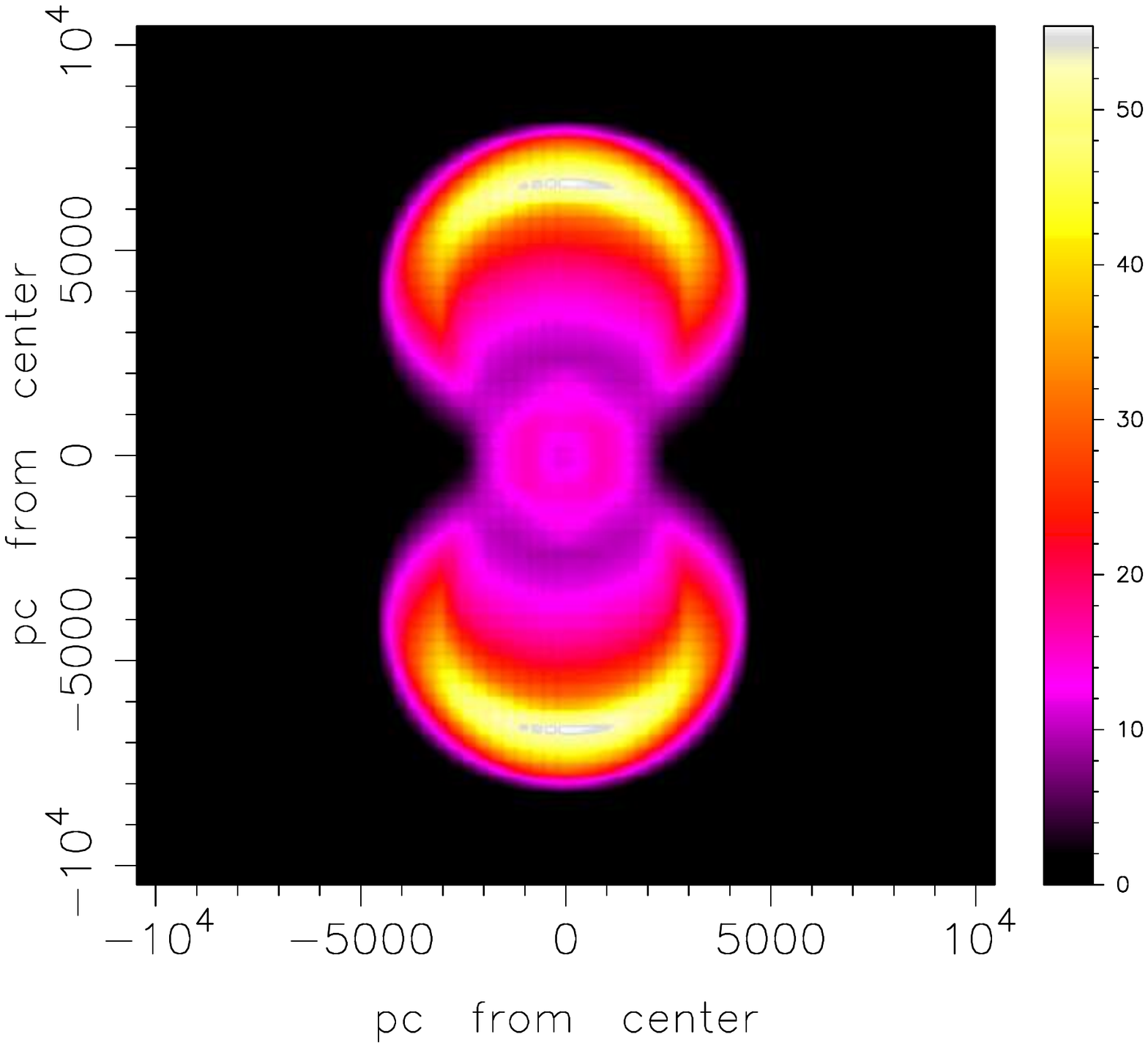}
\caption 
{
Map of the theoretical intensity  of
the Fermi bubbles 
for the    model
in  the presence of 
self-gravity
as
given by Eq.~(\ref{sech2})
with  parameters  as 
in Figure \ref{fermisb_theo_obs_sech2}.
The three Euler angles
characterizing the   orientation
  are $ \Phi $=0$^{\circ }$,
$ \Theta     $=90$^{\circ }$
and   $ \Psi $=90$^{\circ }$.
}
    \label{fermisb_heat_sech2}
    \end{figure}

\begin{figure*}
\begin{center}
\includegraphics[width=7cm]{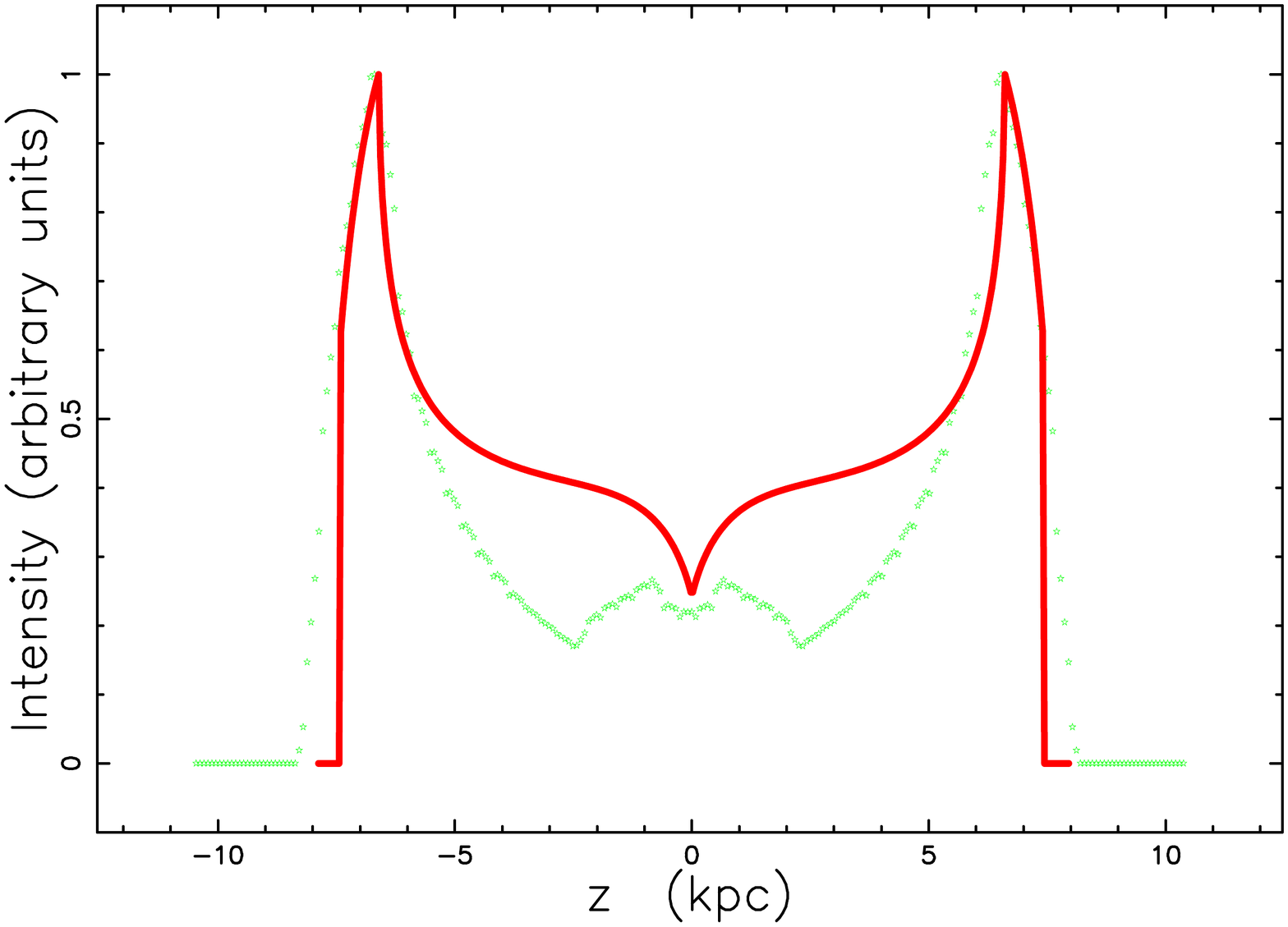}
\end {center}
\caption
{
The intensity profile for   
the Fermi bubbles 
evolving in  a model
in  the presence of 
self-gravity
as
given by Eq.~(\ref{sech2})
along the $z$-axis,
parameters as  in 
Figure  \ref{fermisb_heat_sech2} (green empty stars)
and
the intensity profile for the  piriform model 
along the $z$-axis when  
when $a=3.3$\ kpc,
$b=1.65$\ kpc,
$c=0.825$
and
$C_m$=1 
(red full line).
}
\label{cut_piriform_simula}
    \end{figure*}

The source of the luminosity is assumed here to be
the flux of kinetic energy, $L_m$.
The   observed luminosity along a given direction 
can  be expressed as 
\begin{equation}
L  = \epsilon  L_{m}
\label{luminosity}
\quad  ,
\end{equation}
where  $\epsilon$  is  a constant  of conversion
from  the mechanical luminosity   to  the
observed luminosity, for more details  see \cite{Zaninetti2018c}.
The image of the Fermi bubbles
is shown in Figure \ref{fermisb_heat_sech2}
and  Figure \ref{cut_piriform_simula}
shows a cut of the intensity along the $z$-axis. 

Figure \ref{cut_piriform_simula} also shows 
the cut  of the  piriform  model 
in order to evaluate  the goodness of the
analytical model for complex sections.

\section{Conclusions}

{\bf Equations of motion}
We derived   two equations of motion coupling 
the thin layer approximation with 
the conservation of energy.
The first model implements   a profile
in  the presence of 
self-gravity
of density and the second   a Gaussian profile 
of  density.
In the absence   of analytical  results  for
the trajectory, with the exception  of a Taylor  expansion,
we provided a numerical solution.

{\bf Comparison with other approaches}

As  an example, Figure 3  
in  \cite{Pon2016}   models the  Eridanus--Orion structure  
with an ellipsoid, here we introduce the  mushroom shape, 
see  Figure \ref{fermisb_theo_obs_sech2} relative to the Fermi bubble
and the egg shape, see Figure \ref{w4_3d_gauss}
relative  to W4.
We also  suggested a  first  model   for shapes apparently  impossible to be 
simulated,  see Figure \ref{theo_bump_w4} for the  bump of  W4.

{\bf Theory of the image} 
The introduction  of the piriform curve as a model  for  the 
section of the super-bubble  confirms the existence  
of a  characteristic `U' shape which has  
a maximum in the internal piriform 
at $z=2\,a$ 
and a minimum at the centre, $z=0$, see Eq.~(\ref{cut_piriform}).
The  superposition  of a numerical cut  with the
piriform's cut,
see  Figure \ref{cut_piriform_simula},
shows us that  the use  of the piriform curve 
as a model  is acceptable.

\providecommand{\newblock}{}

\end{document}